\documentclass[aps,prd,twocolumn,superscriptaddress,preprintnumbers,floatfix,nofootinbib,notitlepage,showkeys,showpacs]{revtex4-1}

\usepackage[utf8]{inputenc}
\usepackage{cancel}

\usepackage{graphicx}
\usepackage{hyperref}
\usepackage{latexsym}
\usepackage{amsmath}
\usepackage{amssymb}
\usepackage{bbm}

\usepackage[normalem]{ulem}
\usepackage{pdfsync}
\usepackage{epsfig}
\usepackage{epstopdf}
\usepackage{subfigure}
\usepackage{color}
\usepackage{comment}
\usepackage{slashed}
\usepackage{placeins}
\usepackage{cleveref}

\usepackage{xcolor}
\definecolor{cites}{RGB}{0,180,0}
\definecolor{links}{RGB}{200,0,0}
\hypersetup{colorlinks=true,citecolor=cites,linkcolor=links,urlcolor=blue}

\newcommand{\be}{\begin{equation}} 
\newcommand{\ee}{\end{equation}}
\newcommand{\bea}{\begin{eqnarray}} 
\newcommand{\eea}{\end{eqnarray}}

\newcommand{\bmp}{\noindent\begin{minipage}{16cm}}
\newcommand{\emp}{\end{minipage}\vskip 7mm} 
\def\lsim{\mathrel{\raise.3ex\hbox{$<$\kern-.75em\lower1ex\hbox{$\sim$}}}}
\def\gsim{\mathrel{\raise.3ex\hbox{$>$\kern-.75em\lower1ex\hbox{$\sim$}}}}
\newcommand{\intron}[1]{}

\newcommand{\MSb}{\overline{\textrm{MS}}}

\newcommand{\A}{{a}}
\newcommand{\gGF}{g_{\rm GF}}

\newcommand{\krsout}[1]{}

\begin{document}

\title{Slope of the beta function at the fixed point of SU(2) gauge theory with six or eight flavors}

\author{Viljami Leino}
\email{viljami.leino@helsinki.fi}
\affiliation{Department of Physics, University of Helsinki \\
                      P.O.~Box 64, FI-00014, Helsinki, Finland}
\affiliation{Helsinki Institute of Physics, \\
                      P.O.~Box 64, FI-00014, Helsinki, Finland}

\author{Kari Rummukainen}
\email{kari.rummukainen@helsinki.fi}
\affiliation{Department of Physics, University of Helsinki \\
                      P.O.~Box 64, FI-00014, Helsinki, Finland}
\affiliation{Helsinki Institute of Physics, \\
                      P.O.~Box 64, FI-00014, Helsinki, Finland}

\author{Kimmo Tuominen}
\email{kimmo.i.tuominen@helsinki.fi}
\affiliation{Department of Physics, University of Helsinki \\
            P.O.~Box 64, FI-00014, Helsinki, Finland}
\affiliation{Helsinki Institute of Physics, \\
            P.O.~Box 64, FI-00014, Helsinki, Finland}

\begin{abstract}
We consider measurement of the leading irrelevant scaling exponent $\gamma_g^\ast$, given by the slope of the beta function,
at the fixed point of SU(2) gauge theory
with six or eight flavors. We use the running coupling
measured using the gradient flow method and perform the continuum
extrapolation by interpolating the measured beta function.
We study also the dependence of the results on different
discretization of the flow. For the eight flavor theory we find
$\gamma_g^\ast=0.19(8)_{-0.09}^{+0.21}$. Applying the same analysis
also for the six flavor theory, we find
$\gamma_g^\ast=0.648(97)_{-0.1}^{+0.16}$ consistently
with the earlier analysis.
\end{abstract}

\preprint{HIP-2018-10/TH}

%
\maketitle

\section{Introduction}
One of the basic goals of Beyond Standard Model lattice gauge theory is
to establish the existence of infrared fixed points (IRFP) of
gauge theories with sufficiently large number of flavors and to determine its
properties. For recent reviews
see~\cite{Pica:2017gcb,Nogradi:2016qek,DeGrand:2015zxa}.

A much studied case is SU(2) gauge theory with fermions in the fundamental
representation~\cite{Ohki:2010sr,Bursa:2010xn,Karavirta:2011zg,Hayakawa:2013maa,Appelquist:2013pqa,Leino:2017lpc}.
While the upper edge of the conformal window is robust, as the asymptotic freedom is lost at $N_f=11$,
a consistent picture of the extent of the conformal window has only recently emerged: simulations of the 8 flavor theory have shown the existence of a fixed point~\cite{Leino:2017lpc} and similarly the 6 flavor case~\cite{Leino:2017hgm}.
Theories with $N_f=4$ and $N_f=2$ are expected to break chiral symmetry~\cite{Karavirta:2011zg}. Another benchmark case, where the
existence of a fixed point has been established, is SU(2) gauge theory with two Dirac fermions in the adjoint representation~\cite{Hietanen:2008mr,Hietanen:2009az,
  DelDebbio:2008zf,Catterall:2008qk,
  Bursa:2009we,DelDebbio:2009fd,DelDebbio:2010hx,
  DelDebbio:2010hu,Bursa:2011ru,DeGrand:2011qd, Patella:2012da,
  Giedt:2012rj,DelDebbio:2015byq,
  Rantaharju:2015yva,Rantaharju:2015cne}.

In this paper we analyze further SU(2) gauge theory with eight or six flavors
in the fundamental representation. We use the data generated in~\cite{Leino:2017lpc,Leino:2017hgm}.
For this data the extensive analysis of~\cite{Leino:2017lpc,Leino:2017hgm} demonstrated the
existence of a fixed point and we do not redo this analysis here.
Rather, we focus on the measurement of critical exponent $\gamma_g^\ast$,
given by the slope of the $\beta$-function at the IRFP. For the first time we determine this scheme independent
observable in the eight flavor theory, while the earlier results on the six flavor theory serve as a check on our methodology.

The slope of the $\beta$-function is directly
measurable from the step scaling function of the coupling. We obtain
$\gamma_g^\ast=0.19(8)_{-0.09}^{+0.21}$ in the eight flavor theory, and similar analysis
applied to the six flavor theory yields
$\gamma_g^\ast=0.648(97)_{-0.1}^{+0.16}$, consistent with
the earlier analysis. 

This paper is structured as follows:
We first discuss briefly the lattice implementation:
in Sec.~\ref{sec:model}, and the measurement of the coupling in Sec.~\ref{sec:gfc}.
Then we present our results on the measurement of $\gamma_g^\ast$ for the six and eight flavor theories in Sec.~\ref{sec:results}.
We end with conclusions and outlook in Sec.~\ref{sec:checkout}

\vspace{2mm}
\section{Lattice implementation}
\label{sec:model}
We extend our analysis of the data generated in the studies~\cite{Leino:2017lpc,Leino:2017hgm}.
As the raw data and algorithmic details about the model are available in these papers,
the discussion here will be in a form of brief summary.

The lattice formulations uses HEX-smeared~\cite{Capitani:2006ni} clover improved Wilson fermions
together with gauge action that mixes the smeared and unsmeared gauge actions with mixing parameter $c_g=0.5$:
\begin{displaymath}
	S = (1-c_g)S_G(U) + c_g S_G(V) + S_F(V) + c_{\rm SW} \delta S_{SW}(V) \,,
\end{displaymath}%
where the $V$ and $U$ are the smeared and unsmeared gauge fields respectively.
This mixing of the smeared and unsmeared gauge actions helps to avoid the unphysical bulk phase transition
within the interesting region of the parameter space~\cite{DeGrand:2011vp} enabling simulations at larger couplings.
In the fermion action, we set the Sheikholeslami-Wohlert coefficient to the tree-level value $c_{\rm SW}= 1$,
which is the standard choice for smeared clover fermions~\cite{DeGrand:2011qd,Capitani:2006ni,Shamir:2010cq}.
In earlier studies~\cite{Karavirta:2011zg,Karavirta:2011mv} we have verified
that this value is very close to the true non-perturbatively fixed $c_{\rm SW}$ coefficient and cancels most of the $\mathcal{O}(a)$ errors.

We use Dirichlet boundary conditions at the temporal boundaries $x_0=0$, $L$,
as in the Schrödinger functional method~\cite{Luscher:1992an,Luscher:1992ny,Luscher:1993gh,DellaMorte:2004bc},
by setting fermion fields to zero and gauge link matrices to unity $U=V=1$.
The spatial boundaries are periodic.
These boundary conditions allow us to tune the fermion mass to zero using the PCAC relation.
In practice, the hopping parameter $\kappa_c(\beta_L)$ is tuned at lattices of size $24^4$,
so that the PCAC fermion mass vanishes with accuracy $10^{-5}$.
This critical hopping parameter $\kappa_c(\beta_L)$ is then used on all the lattice sizes.

The simulations are run using the hybrid Monte Carlo algorithm with 2nd order Omelyan integrator~\cite{Omelyan:2003:SAI,Takaishi:2005tz}
and chronological initial values for the fermion matrix inversions~\cite{Brower:1995vx}.
We reach acceptance rate that is larger than 85\%.
For the analysis considered in this paper, we use lattices of volumes $(L/a)^4=10^4,12^4,16^4,18^4,20^4,24^4,30^4,32^4$,
where the $L=18,30$ are only used for $N_f=6$ results and $L=32$ is only available in the $N_f=8$ analysis.
The difference in available lattice sizes between the two cases is caused by the fact that we used step scaling step
$s=2$ for $N_f=8$ in~\cite{Leino:2017lpc} and $s=3/2$ for $N_f=6$ in~\cite{Leino:2017hgm}.
The bare couplings $\beta_L \equiv 4/g_0^2$ vary from 8 to 0.5 for $N_f=6$ and to 0.4 for $N_f=8$.
For all combinations of $L/a$ and $\beta_L$, we generate between $(5-100)\cdot 10^{3}$ trajectories.

\section{Gradient flow coupling constant}
\label{sec:gfc}
The running coupling is defined by the Yang-Mills gradient flow method~\cite{Narayanan:2006rf,Luscher:2009eq,Ramos:2015dla}.
In the lattice flow equation the unsmeared lattice link variable $U$
is evolved using the tree-level improved L\"uscher-Weisz pure gauge action.

The coupling at scale $\mu=1/\sqrt{8t}$~\cite{Luscher:2010iy} is defined via the energy measurement as
\begin{align}
	\label{eq:g2gf}
	\gGF^2(\mu) &= \mathcal{N}^{-1}t^2 \langle E(t) \rangle\vert_{x_0=L/2\,,\,t=1/8\mu^2}\,,
\end{align}%
where $a$ is the lattice spacing. The renormalization factor $\mathcal{N}$ has been calculated in Ref.~\cite{Fritzsch:2013je}
for the Schrödinger functional boundary conditions so that $\gGF^2$ matches continuum $\MSb$ coupling in the tree level of perturbation theory.
Since the Schrödinger functional boundary conditions break the translation invariance in time direction,
we measure the coupling only at central time slice $x_0=L/2$.
We measure the energy density $E(t)$ using both the clover and plaquette discretizations.

The flow time $t$ at which the gradient flow coupling is evaluated is arbitrary and defines the renormalization scheme.
However, it is useful to link the lattice and renormalization scales with a dimensionless parameter $c_t$ so
that the relation $\mu^{-1} = c_tL = \sqrt{8t}$ is satisfied~\cite{Fodor:2012td,Fritzsch:2013je}.
In Ref.~\cite{Fritzsch:2013je} it is proposed, that for the Schrödinger functional boundary conditions
the choice $c_t=0.3-0.5$ yields reasonably small statistical variance and cutoff effects.
For both the eight~\cite{Leino:2017lpc} and six flavor cases~\cite{Leino:2017hgm} we did full analysis within this
range of $c_t$ and found universal behavior compatible with the existence of a fixed point independently of the value
of $c_t$. Since we know from these previous studies~\cite{Leino:2017lpc,Leino:2017hgm} that the $c_t$ has a little effect
on the measurement of the scheme independent quantities, we use the same choices for $c_t$ in this study as we
reported our final results in Refs.~\cite{Leino:2017lpc,Leino:2017hgm}. 
We choose $c_t=0.3$ for the six flavor theory and $c_t=0.4$ for the eight flavor theory.

In the earlier studies~\cite{Leino:2017lpc,Leino:2017hgm},
we reduced the $\mathcal{O}(a^2)$ discretization effects by using the $\tau_0$-correction method~\cite{Cheng:2014jba},
that modifies the Eq.~\eqref{eq:g2gf} by measuring the energy density at flow time $E(t+\tau_0 a^2)$.
The $\tau_0$ was tuned by hand to remove most order $\mathcal{O}(a^2)$ effects.
Since the discretization effects grow with the coupling, we made the $\tau_0$ function of the gradient flow coupling $\gGF^2$; see~\cite{Leino:2017lpc,Leino:2017hgm} for the details of this implementation.

In the present work we would like to investigate an alternative method to reduce the $\mathcal{O}(a^2)$ correction.
In Ref.~\cite{Fodor:2015baa} it is noted that as the different discretizations have different $\mathcal{O}(a^2)$ behavior,
it is possible to combine two discretizations so that the $\mathcal{O}(a^2)$ effects cancel each other.
Combining the gradient flow coupling measurements done with the plaquette and clover discretizations, we therefore get
\begin{equation}
\label{eq:alternative_a2}
\gGF^2=\mathcal{N}^{-1}t^2\left[ (1-X)\langle E_\mathrm{Clover}(t)\rangle + X\langle E_\mathrm{Plaq.}(t)\rangle\right]\,,
\end{equation}
where mixing coefficient $X$ can in principle be chosen freely, but the perturbative results
for periodic boundaries from  Refs.~\cite{Fodor:2014cpa,Kamata:2016any} suggest value of $X=1.25$ for our choice of discretizations.  We will investigate the dependence of the results on the value of the mixing
parameter $X$.

Since the $\tau_0$-correction was optimized for the whole data and depended on the measured coupling,
it naturally gives more fine tuned correction.
On the other hand, in this paper we are interested only on the quantities at the IRFP so the data does not
have to be perfectly improved at small couplings.
Also, we will do bulk of our analysis with the unimproved $X=0$ and then only use the parameter $X$ to
study how the different discretizations affect the results.

\section{Leading irrelevant critical exponent}
\label{sec:results}
Since, based on earlier results of~\cite{Leino:2017lpc,Leino:2017hgm}, 
we know that the lattice configurations we have available for our
analysis imply the existence of a fixed point, we now turn to the details of the analysis relevant for the present work.
We will use two different methods to extract the leading critical exponent: First, we determine the slope of the beta function
directly from the results on the step scaling function in six and eight flavor theory. Second, we apply the finite scaling method
developed in Refs.~\cite{Appelquist:2009ty,DeGrand:2009mt,Lin:2015zpa,Hasenfratz:2016dou}. The first method is robust, while the
second method is more uncertain as it can be applied only in the vicinity of the fixed point whose location must be known
from the outset. Since we know the location of the fixed point, the second method can be applied, but it only serves as a consistency
check of the more robust results of the first method.

\subsection{Step scaling method}

\begin{figure}[t]
  \includegraphics[width=8.6cm]{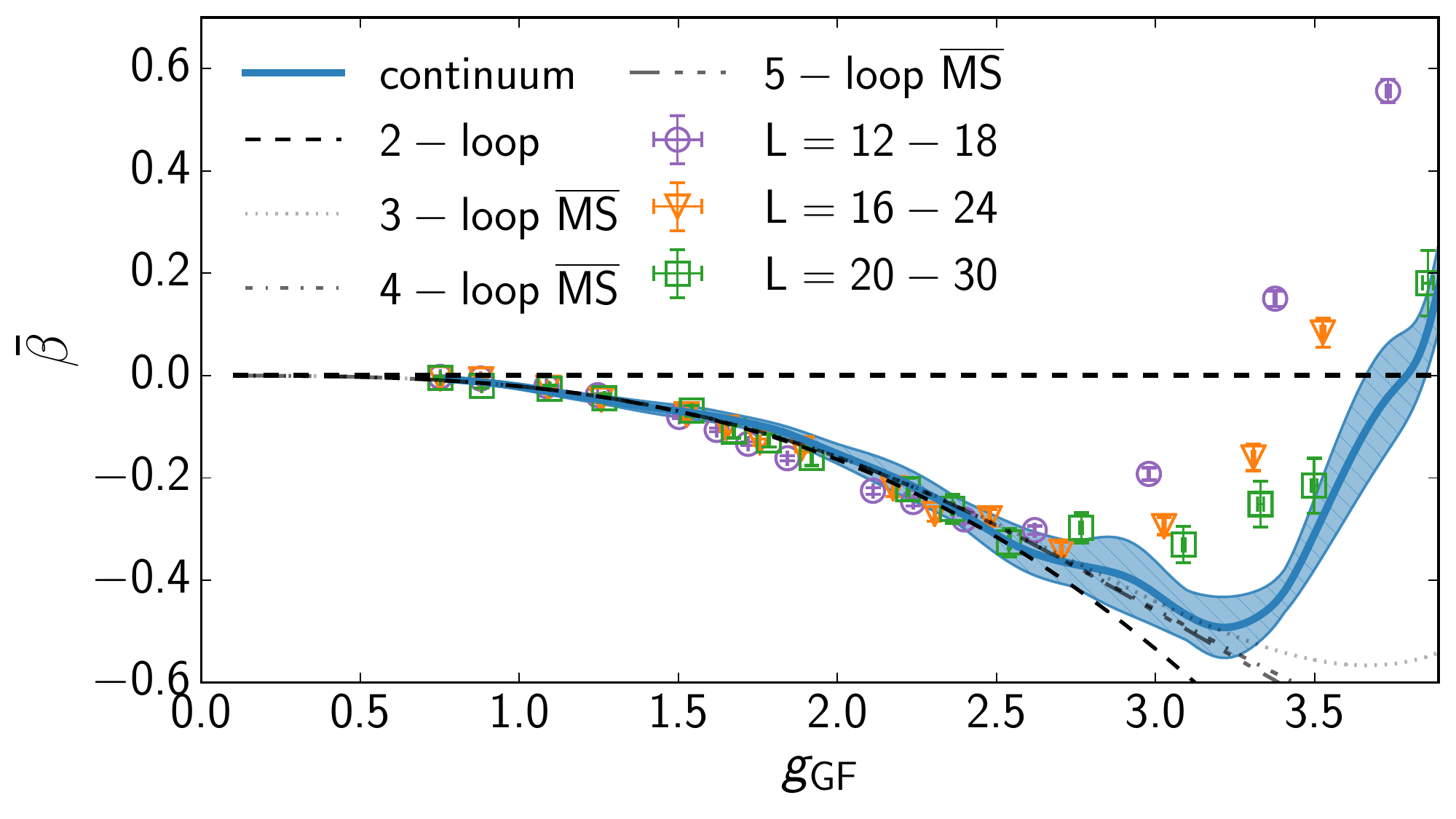}
  \caption[b]{The continuum extrapolated $\beta$-function of the six flavor theory. The gradient flow coupling has been measured
  with $c_t=0.3$ and clover discretization. The $\tau_0$ improvement has been used to reduce the $\mathcal{O}(a^2)$ errors.
  }
\label{fig:nf6old}
\end{figure}%

The leading irrelevant exponent of the coupling $\gamma_g$ is defined as the slope of the $\beta$-function.
On the lattice the evolution of the coupling is measured with the step scaling function
\begin{align}
  \Sigma(g^2,L/\A,s) &= \left . \gGF^2(g^2_0,sL/\A) \right|_{\gGF^2(g^2_0,L/\A)=g^2}, \\
  \sigma(g^2,s) &= \lim_{L/a \rightarrow \infty} \Sigma(g^2,L/a,s)\,.
\end{align}
In the vicinity of the IRFP, where $\beta$-function is small,
the step scaling function, $\beta$-function and $\gamma_g^\ast$ can be related as follows:
\begin{align}
\beta(g) &= \mu \frac{dg}{d\mu} \approx \gamma_g^\ast(g - g^\ast) \label{eq:beta*}\\
&\approx \bar\beta(g) \equiv
\frac{g}{2\ln(s)} \left ( 1 - \frac{\sigma(g^2,s)}{g^2} \right )\,. \label{eq:raw_betaf}
\end{align}%
Here $g^\ast$ is the coupling at the IRFP.
In Ref.~\cite{Leino:2017hgm}
we calculated the step scaling function for $N_f=6$ by interpolating the measured couplings with 9th order polynomial,
which led to continuum extrapolation shown in Fig.~\ref{fig:nf6old} for $c_t=0.3$.
In this case the final form of the function near the fixed point
was smooth enough that we managed to measure the leading irrelevant exponent
$\gamma_g^\ast=0.648(97)_{-0.1}^{+0.16}$, where the first set of errors implies
the statistical errors with the parameters used in Ref.~\cite{Leino:2017hgm},
and the second set of errors gives the variance between all measured discretizations.
When the values of $c_t$ were varied, the $\gamma_g^\ast$ measurements remained
consistent with each other, within the errors, indicating the scheme
independence of this quantity.
In Fig.~\ref{fig:nf6old} we also present the perturbative $\MSb$ results up to 5-loop level~\cite{Herzog:2017ohr}. 
Only the 2-loop result is scheme independent and rest of the curves are shown as a reference. 
As the 5-loop $\MSb$ does not feature an IRFP and evolves mostly outside the figure, we will not plot it in any future figures.

\begin{figure}[t]
  \includegraphics[width=8.6cm]{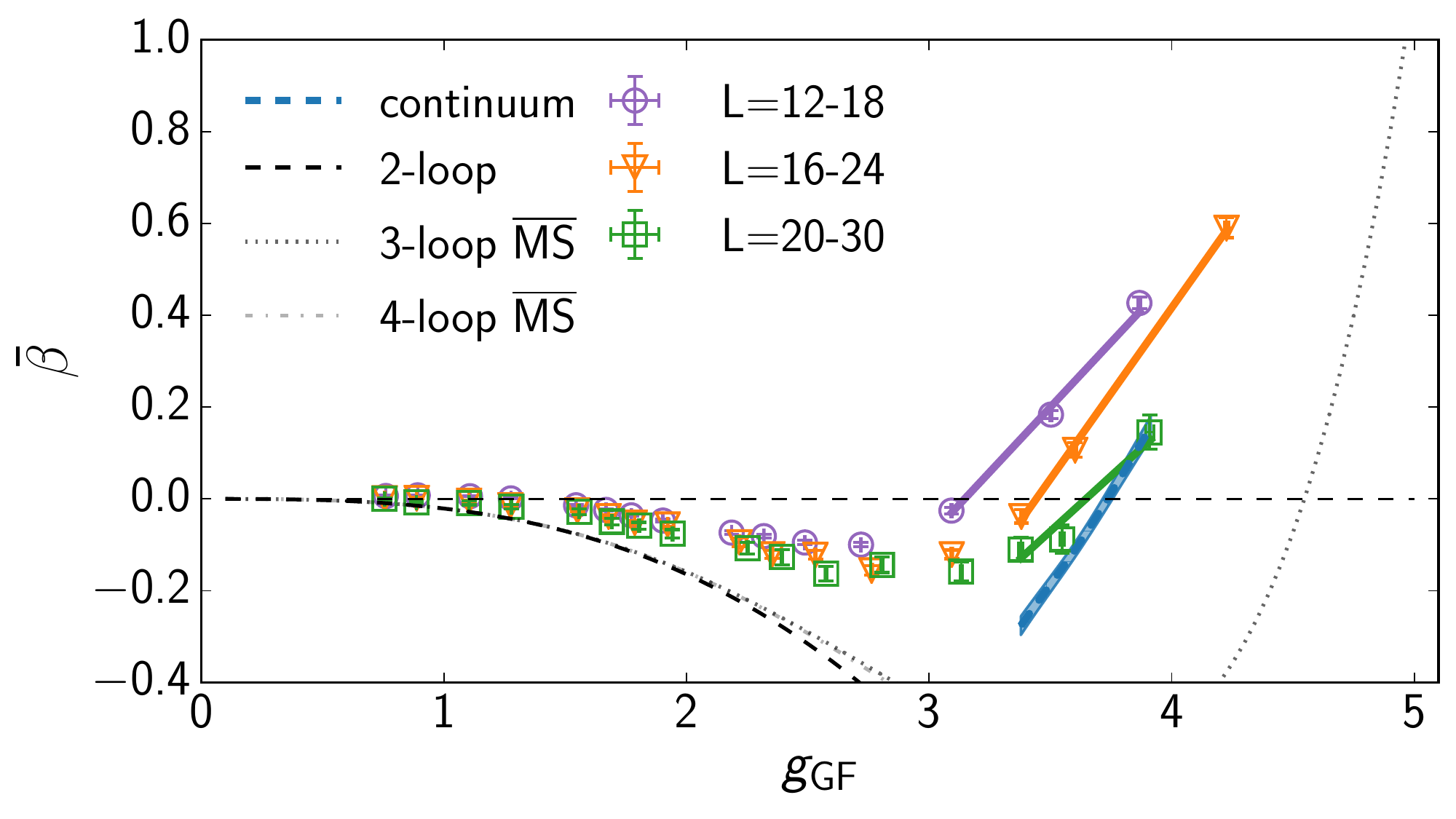}
  \includegraphics[width=8.6cm]{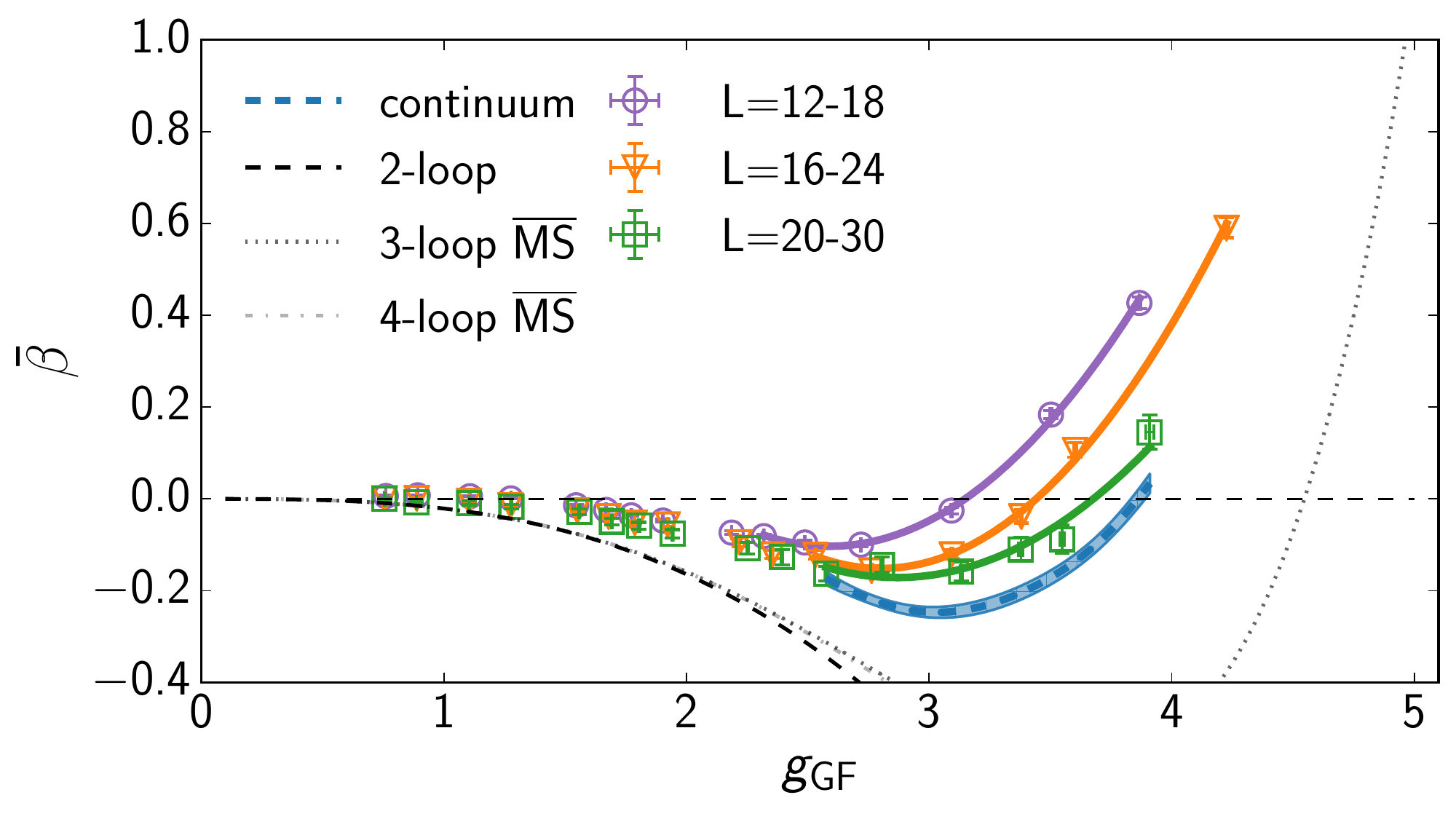}
  \includegraphics[width=8.6cm]{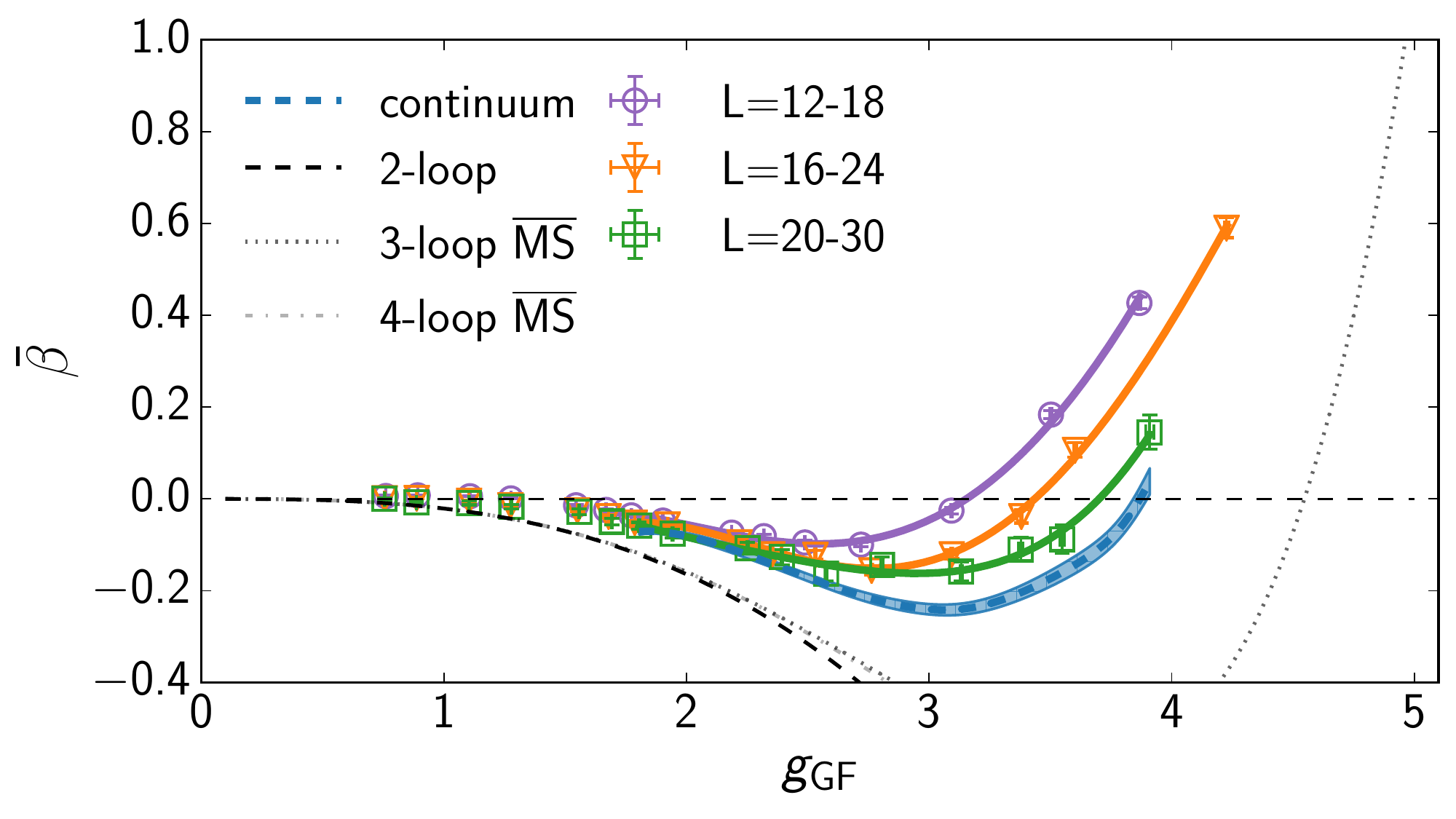}
  \caption[b]{
  			   {\em From top to bottom:} The $N_f=6$ linear, quadratic, and quartic fits to associated ranges of data
			   with $X=0$ and $c_t=0.3$.
  }
\label{fig:nf6linear3}
\end{figure}

However, we can also directly interpolate the finite volume $\bar\beta(g)$-function~\eqref{eq:raw_betaf}
(where $\sigma(g^2,s)$ is substituted with $\Sigma(g^2,L/a,s)$),
instead of the measured couplings.
Similar ideas have previously been implemented in Refs.~\cite{Rantaharju:2015yva,Rantaharju:2015cne,DallaBrida:2016kgh}.
Not only does this make the continuum limit smoother around the fixed point, but allows to limit the fit to a region near the IRFP.
We show three different fits in Fig.~\ref{fig:nf6linear3} for $c_t=0.3$ and $X=0$,
which corresponds to the unimproved clover measurements in~\cite{Leino:2017hgm}.
We use three different polynomial fit functions: linear, quadratic, and quartic,
and for each fit choose the number of points that minimizes the $\chi^2/\mathrm{d.o.f}$.
While Fig.~\ref{fig:nf6linear3} shows the $X=0$ -case, we do similar fits
for $-1.5 \le X \le 1$.  Depending on the value of $X$ the $\chi^2/\mathrm{d.o.f}$
value varies between $0.5$ and $2.5$.

\begin{figure}[t]
  \includegraphics[width=8.6cm]{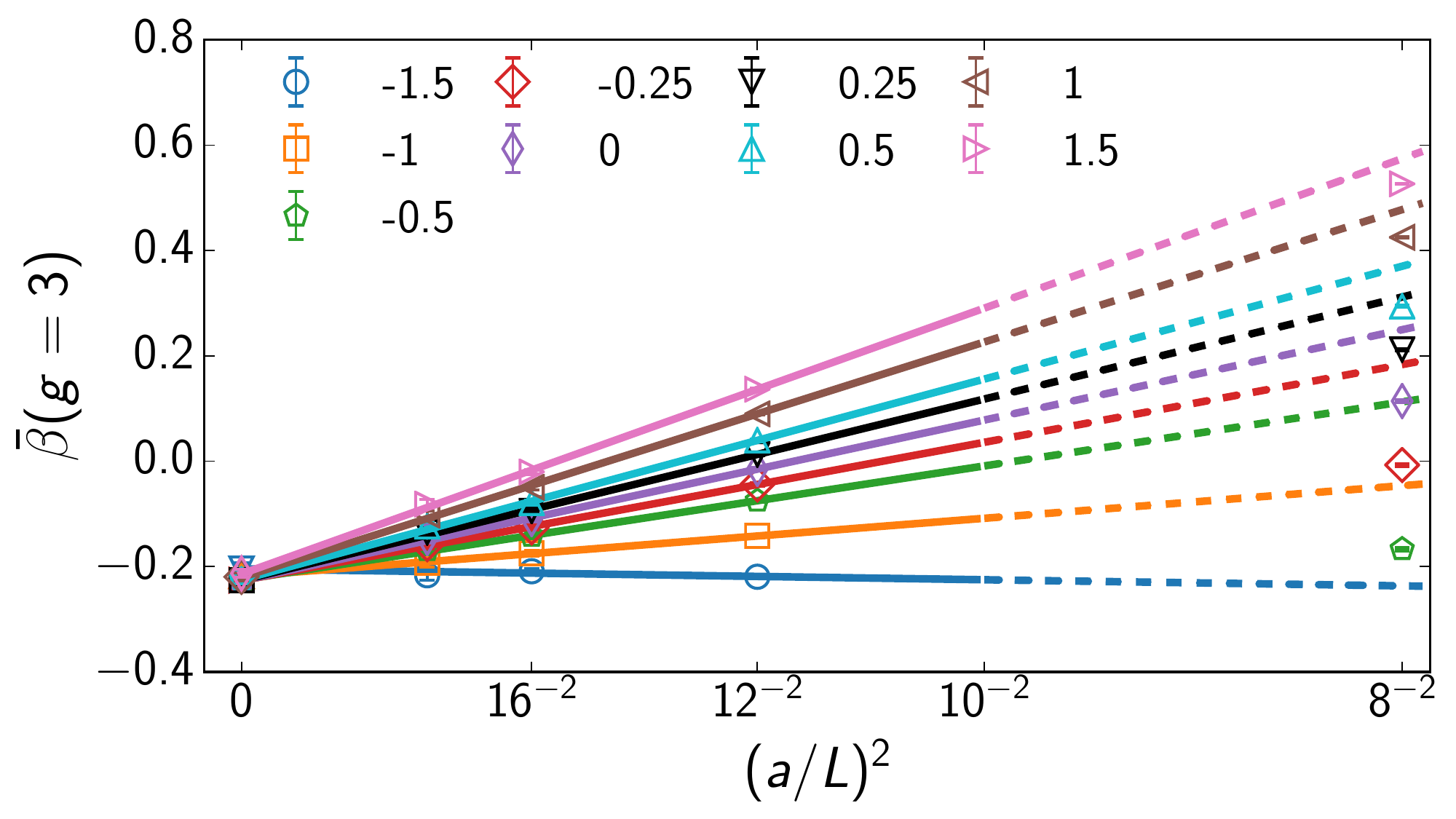}
  \includegraphics[width=8.6cm]{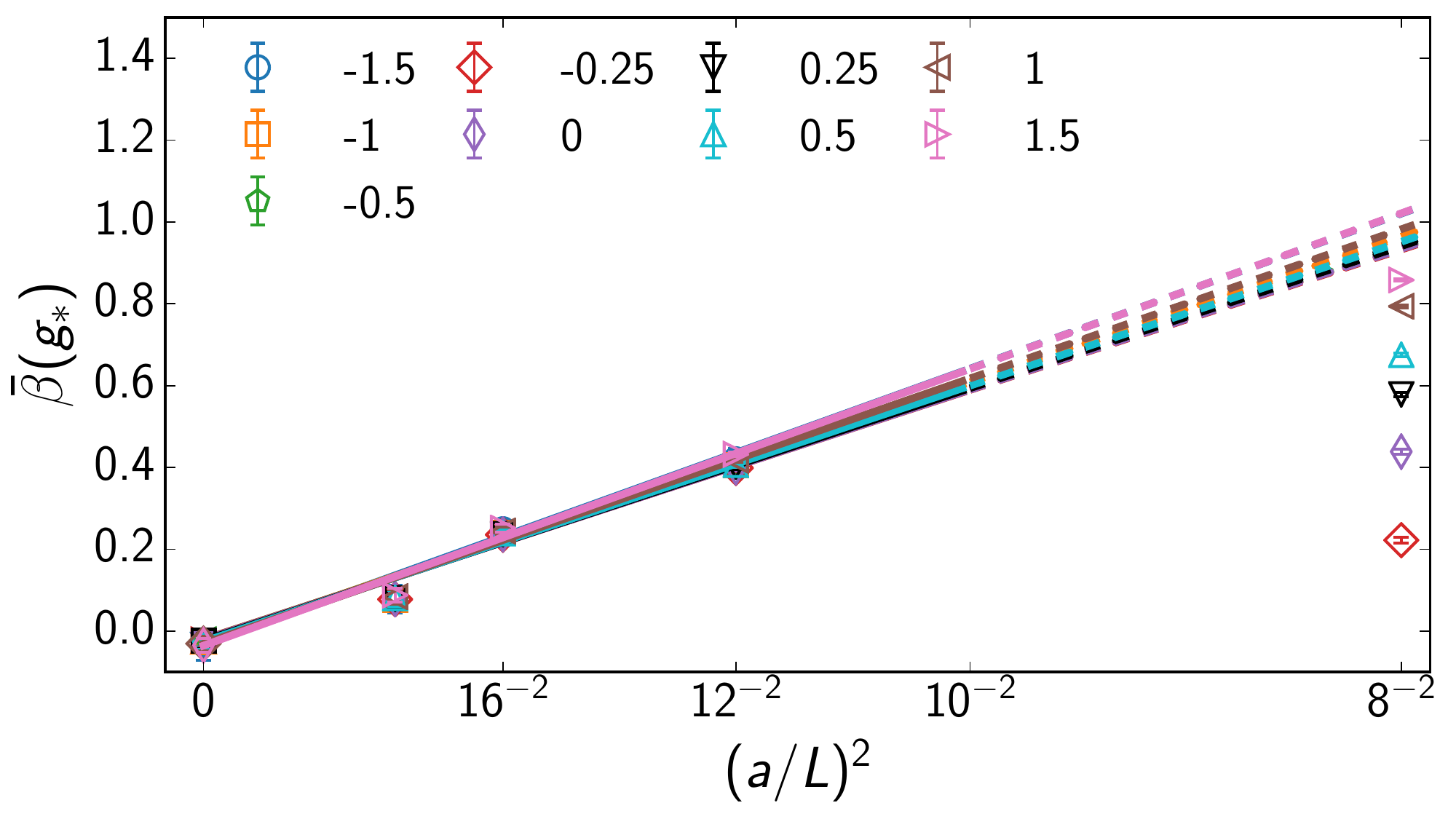}
  \caption[b]{The continuum limit of the $\beta$-function at $g=3$ ({\em top}) and 
  			  in the vicinity of the IRFP $g_\ast=3.8$ ({\em bottom});
              in six flavor theory,
              for different choices of the discretization mixing parameter $X$ using the quadratic fit.
  }
\label{fig:x6_values_comp}
\end{figure}%

In Fig.~\ref{fig:x6_values_comp} we show the continuum limit of $\bar\beta(g_\ast)$
at different values of the mixing parameter $X$, assuming discretization errors ${\cal O}((a/L)^2)$. 
The top and bottom panels correspond, respectively, to the values $\gGF=3$ and $\gGF\approx g^\ast\approx3.8$ of the coupling.
For $\gGF=3$ the $\chi^2/\mathrm{d.o.f}$ varies between 0.5 and 2 depending on the $X$ and for the $\gGF=3.8$
the $\chi^2/\mathrm{d.o.f}$  varies between 2 and 4 depending on the $X$.

From the figure we can see that the continuum limit remains stable with respect to the variations of the parameter $X$.
At weaker coupling the values below $X=-0.5$ have reduced $a^2$-effects, while
near the fixed point the dependence on $X$ becomes less pronounced.

\begin{figure}[t]
  \includegraphics[width=8.6cm]{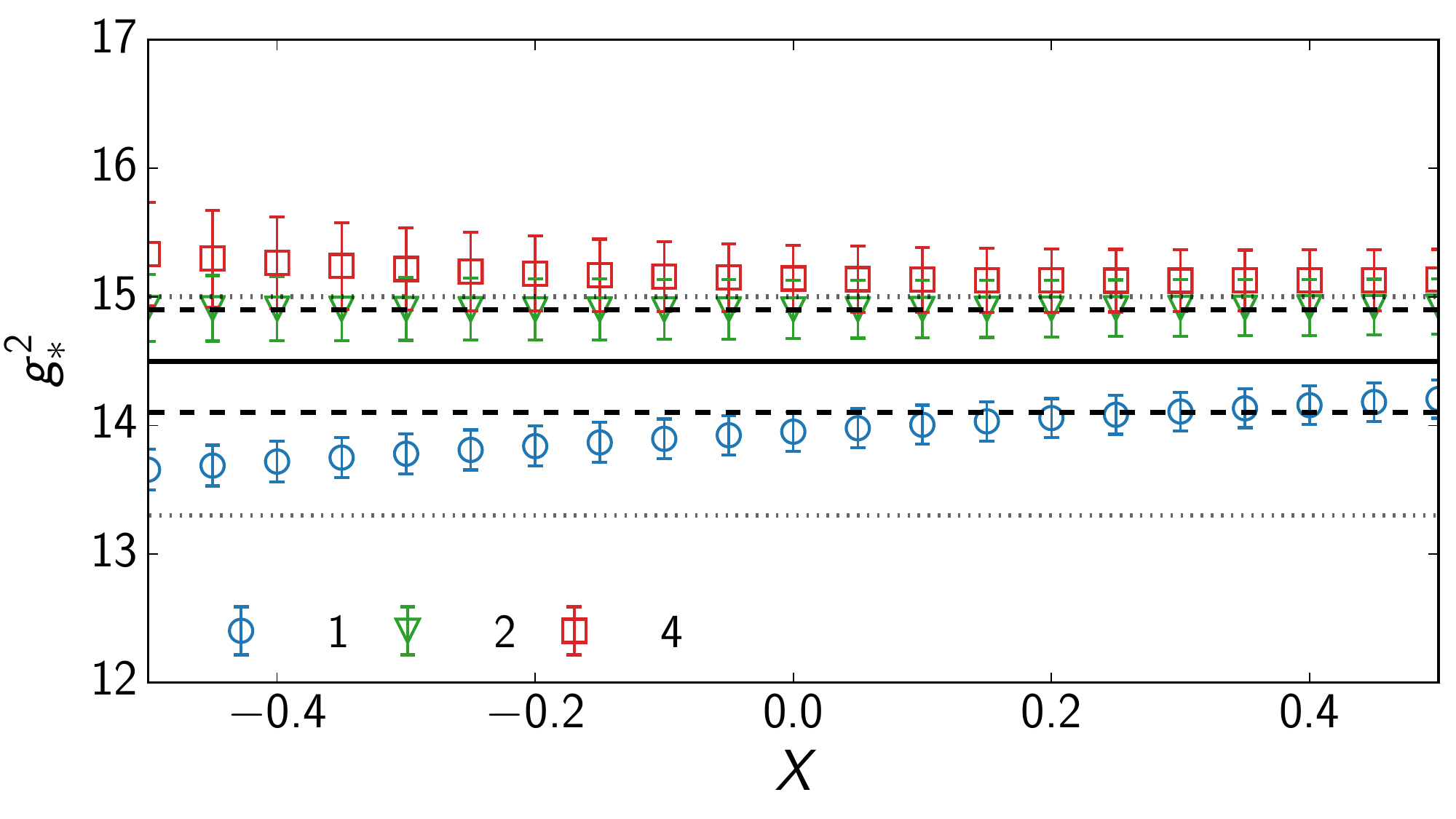}
  \caption[b]{Location of the IRFP as a function of the mixing parameter $X$ at for linear (circles), quadratic (triangles) and quartic (squares) fits.
			  The black lines show the reported result from~\cite{Leino:2017hgm} with its statistical errors, and the gray dotted lines show the variance between different discretizations in~\cite{Leino:2017hgm}.
  }
\label{fig:nf6_mixgpar_aff_g2}
\end{figure}

In Fig.~\ref{fig:nf6_mixgpar_aff_g2} we show the location of the
IRFP as a function of the mixing parameter $X$ and for different fits.
The existence and the location of the IRFP in $N_f=6$ theory
agrees with our previous measurement~\cite{Leino:2017hgm},
$g^2_\ast=14.5(3)^{+0.41}_{-1.38}$,
within the error bounds indicating the variance between different discretizations shown 
with the dotted horizontal lines in the figure 
and corresponding to the second set of errors in the numerical result quoted above.
While all the chosen interpolations agree with the previous measurement,
we note that the linear fit seems to have stronger $X$ dependence than higher order polynomials.
This is most likely caused by the sparsity of the points around the fixed point. 
Since the quadratic fit seems to give most consistent results with previous measurement,
has small $X$ dependence, and has smaller errors than quartic fit, we choose it as our main result.
As the $X$ dependence seems to be small, we will use $X=0$ as our default choice.

\begin{figure}[t]
  \includegraphics[width=8.6cm]{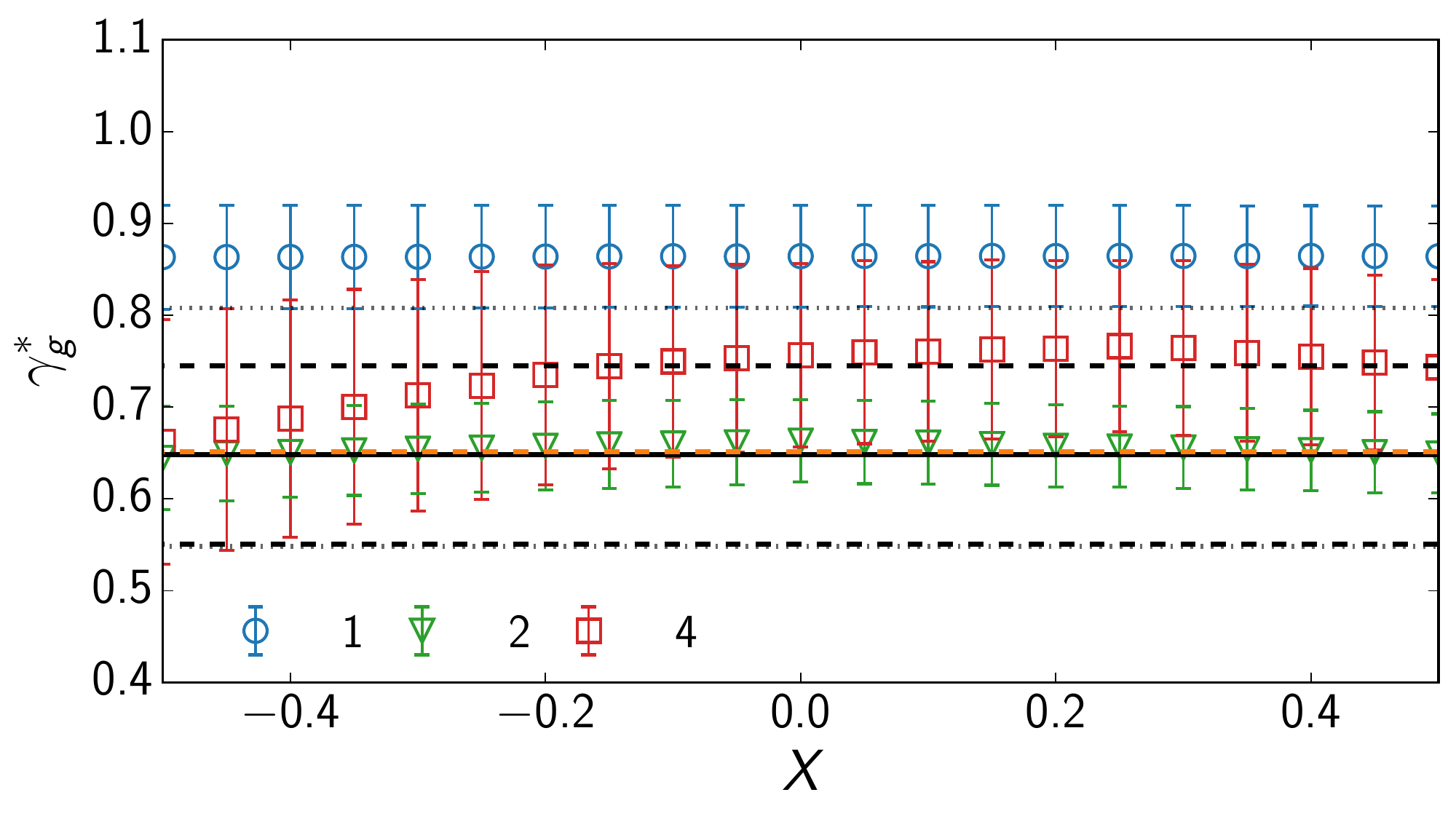}
  \caption[b]{The scaling exponent $\gamma_g^\ast$ for the six flavor theory with linear (circles), quadratic (triangles), and quartic (squares) fits to different ranges of data
  in the vicinity of IRFP.
  The black lines show the reported result from~\cite{Leino:2017hgm} with its statistical errors,
  and the gray dotted lines show the variance between different discretizations in~\cite{Leino:2017hgm}.
  The orange dashed line shows the scheme independent large-$N_f$ perturbative result~\cite{Ryttov:2017kmx,Ryttov:2017toz}.
  }
\label{fig:nf6_linear_comp}
\end{figure}

For the $\gamma_g^\ast$ measurement, 
we reproduce the value of $\gamma^\ast$ obtained in the original analysis in~\cite{Leino:2017hgm}.
Similarly, the results of $\gamma_g^\ast$ measurements are shown in Fig.~\ref{fig:nf6_linear_comp}.
Using the quadratic fit with $X=0$ we get $\gamma_g^\ast=0.66(4)^{+0.25}_{-0.13}$,
where the second set of errors include the variance in both $X$ and between different interpolation functions.
This is in agreement with the result $\gamma_g^\ast=0.648(97)_{-0.1}^{+0.16}$
obtained earlier in~\cite{Leino:2017hgm}.

\begin{figure}[t]
  \includegraphics[width=8.6cm]{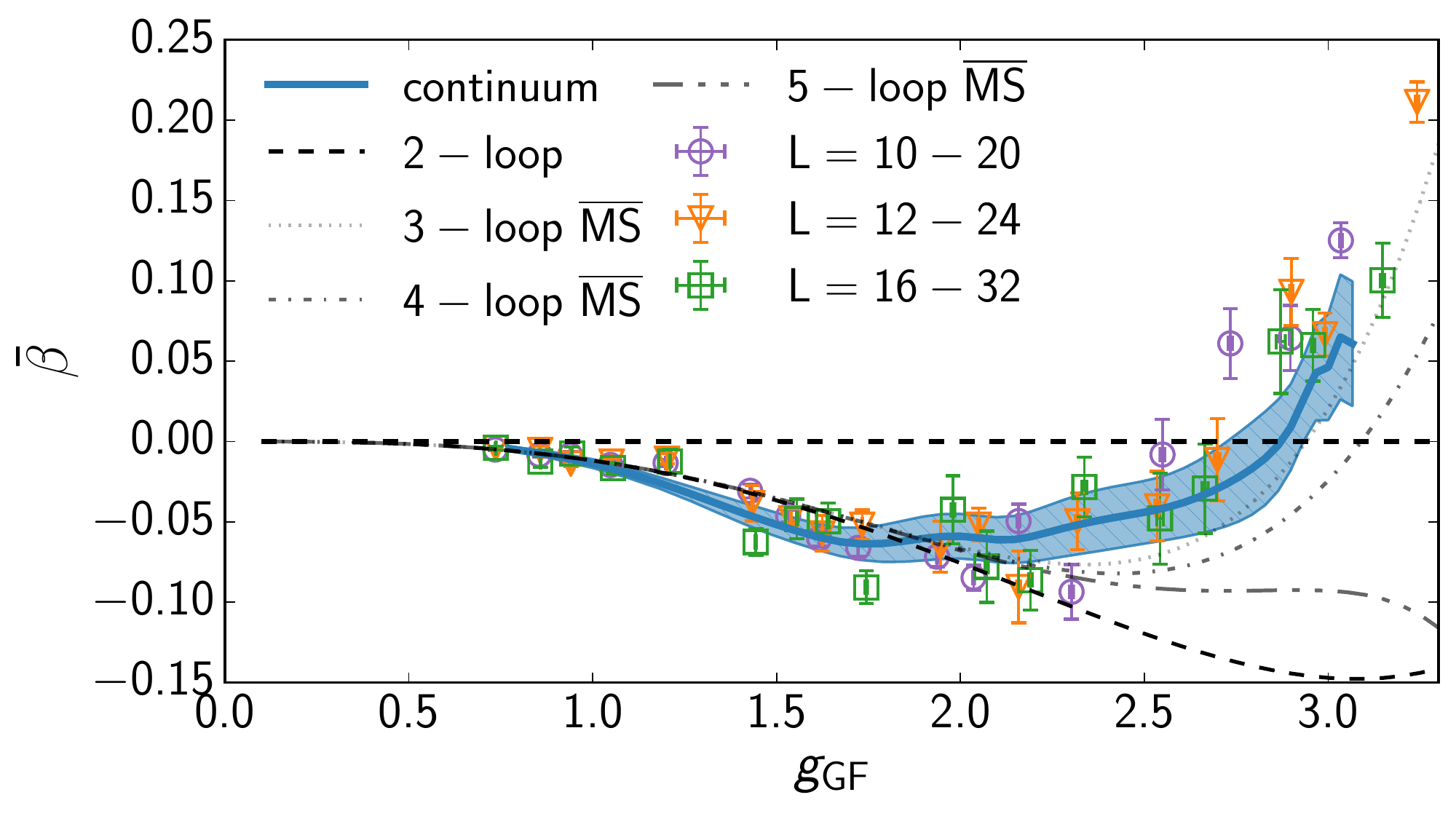}
  \caption[b]{The continuum extrapolated $\beta$-function of the eight flavor theory. The gradient flow coupling has been measured
  with $c_t=0.4$ and clover discretization. The $\tau_0$ improvement has been used to reduce the $\mathcal{O}(a^2)$ errors.
  }
\label{fig:nf8old}
\end{figure}

In Ref.~\cite{Leino:2017lpc} we measured the running coupling for $N_f=8$
by interpolating the couplings with rational ansatz, where the numerator was 7th order polynomial
and the denominator a 1st order polynomial, and then taking the continuum limit.
This continuum limit, together with the raw $\tau_0$-corrected data, is shown in the Fig.~\ref{fig:nf8old}.
The interpolation function was chosen by extensive statistical tests to give best fit to the whole data.
While we were sure to check the existence of the IRFP within the reported errors,
the chosen fit function develops a curvature at the fixed point,
which renders a reliable measurement of $\gamma_g^\ast$ impossible.
Again, we show the $\MSb$ results up to 5-loop order, but will drop the 5-loop curve from future figures, 
as the theory doesn't have an IRFP at this level of loop expansion.

\begin{figure}[t]
  \includegraphics[width=8.6cm]{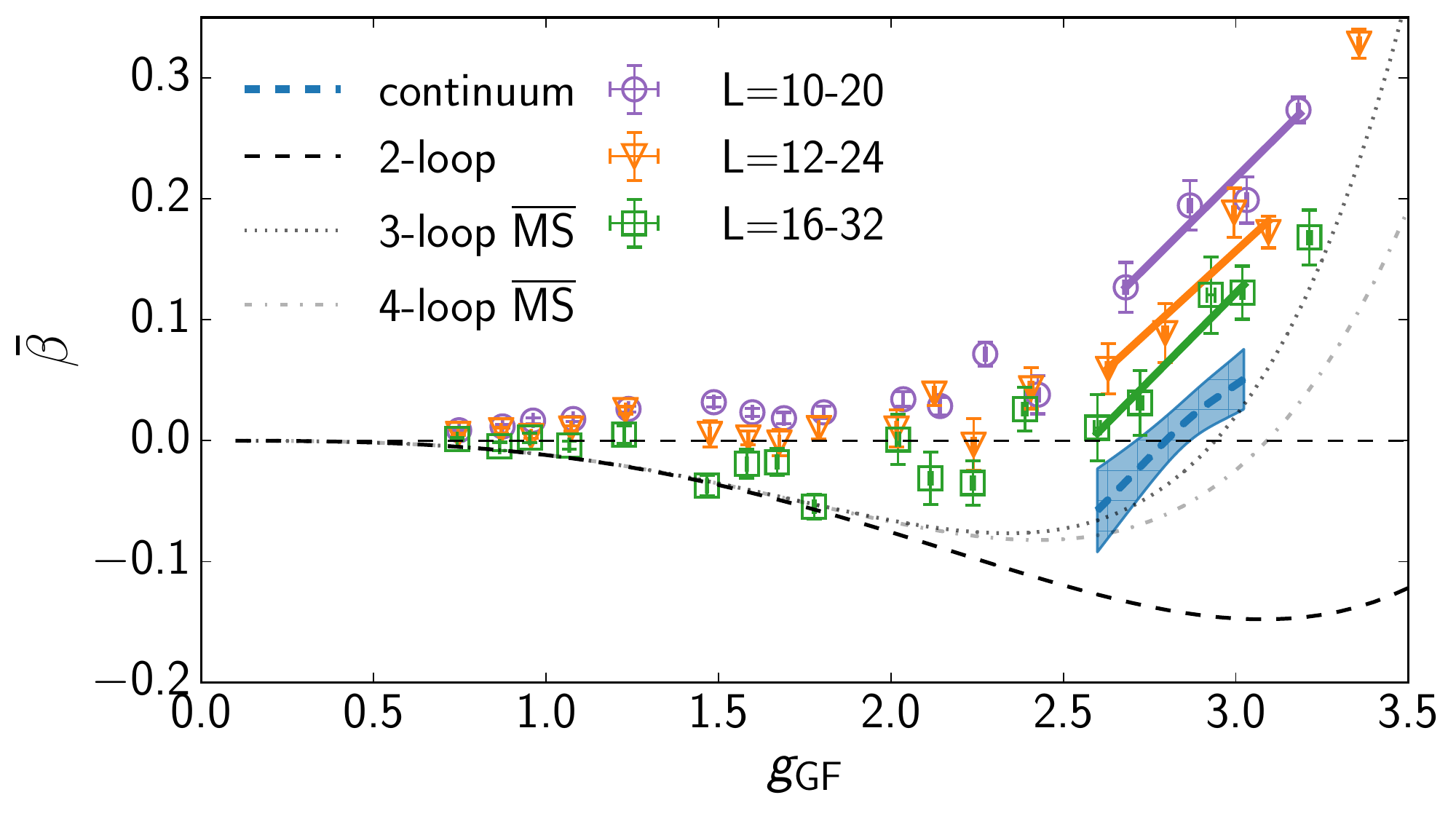}
  \includegraphics[width=8.6cm]{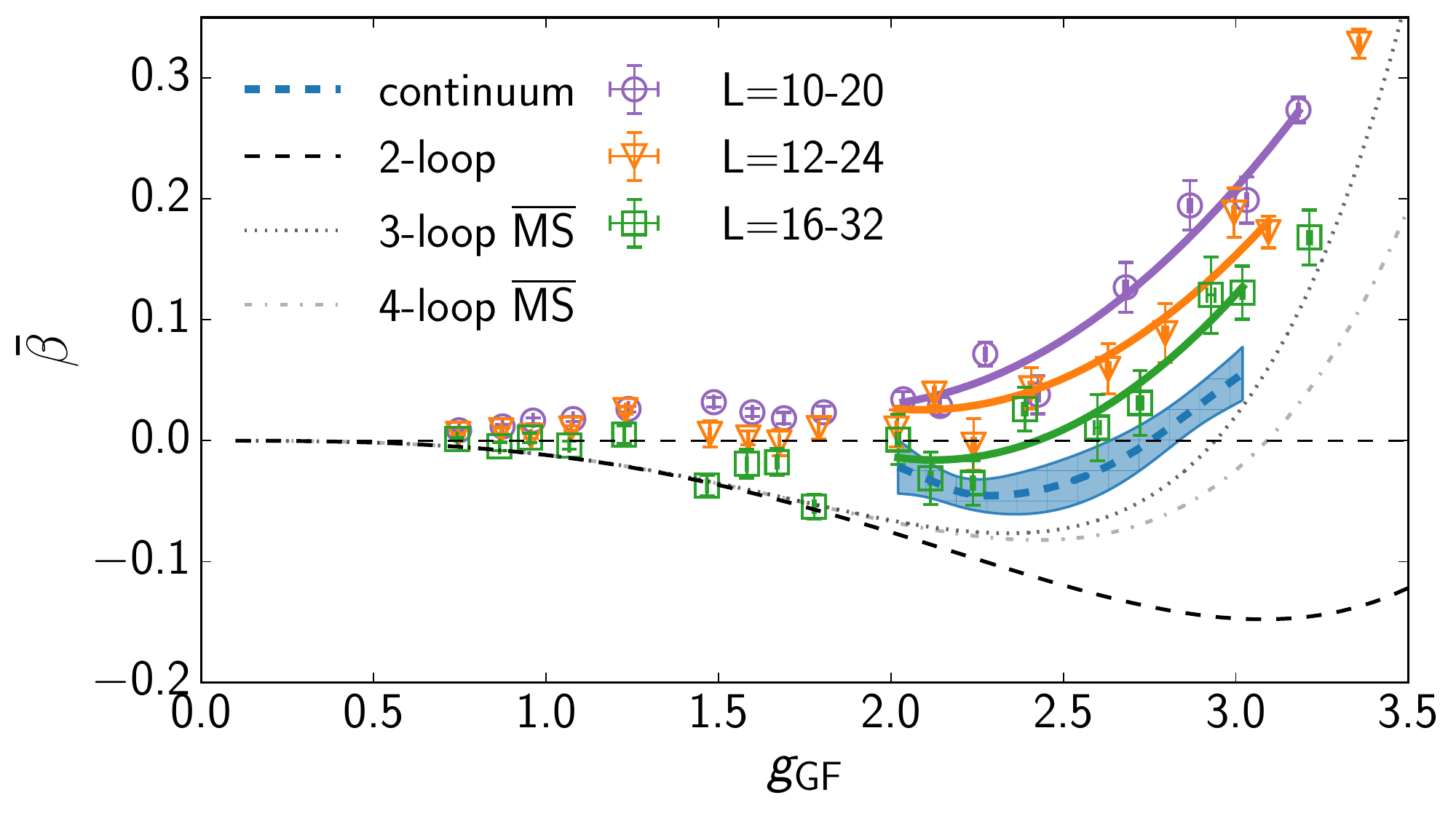}
  \includegraphics[width=8.6cm]{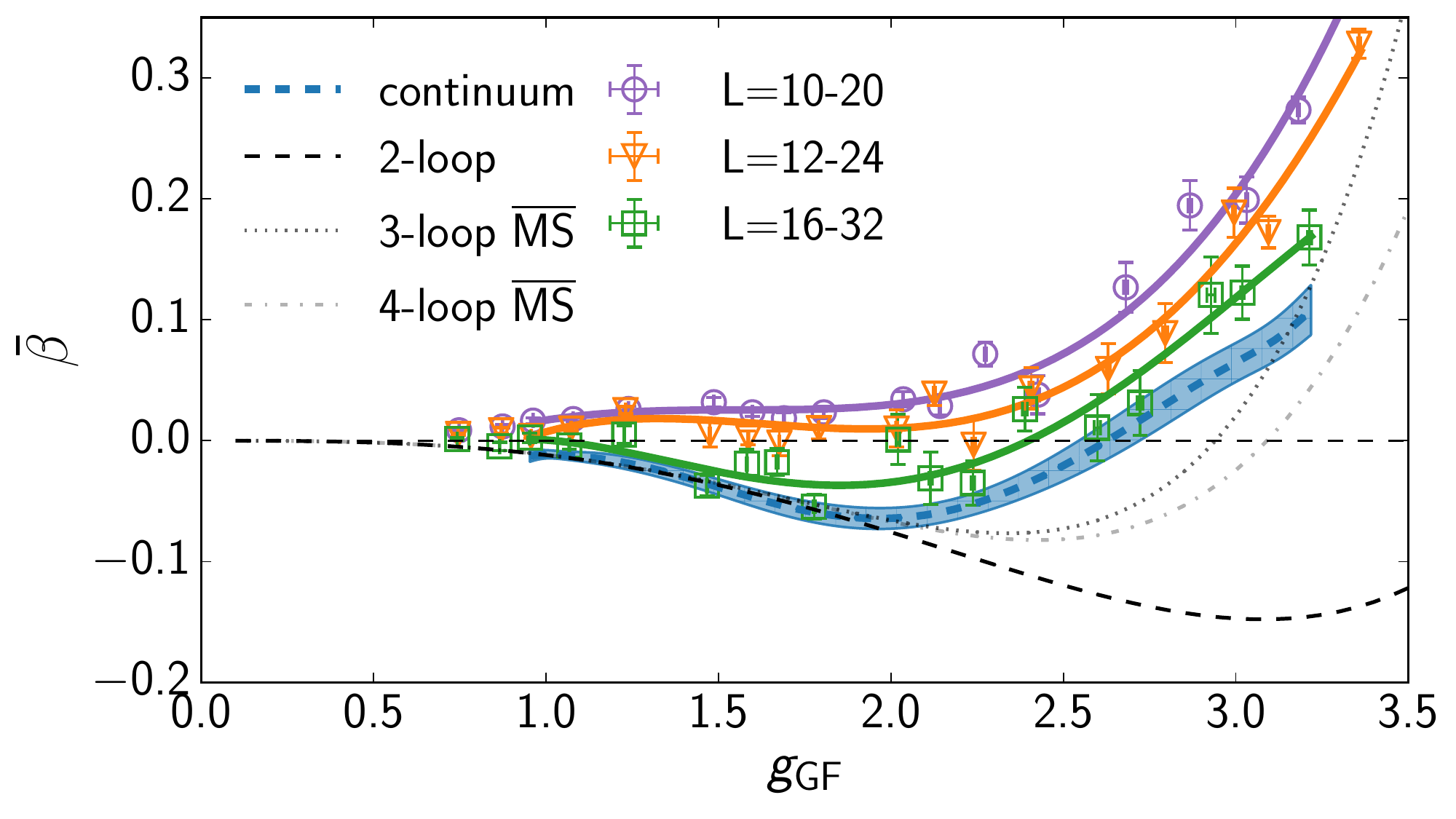}
  \caption[b]{
  			   {\em From top to bottom:} The $N_f=8$ linear, quadratic, and quartic fits to associated ranges of data
			   with $X=0$ and $c_t=0.4$.
  }
\label{fig:nf8linear}
\end{figure}

On the basis of the results in six flavor theory,
we now directly interpolate the raw $\beta$-function instead of the raw couplings also in the $N_f=8$ case.
Again we perform the linear, quadratic, and quartic fits for the regions of data where these fit ansatz give the
best $\chi^2/\mathrm{d.o.f}$. These fits, together with their continuum limits, are shown in Fig.~\ref{fig:nf8linear}.

\begin{figure}[ht]
  \includegraphics[width=8.6cm]{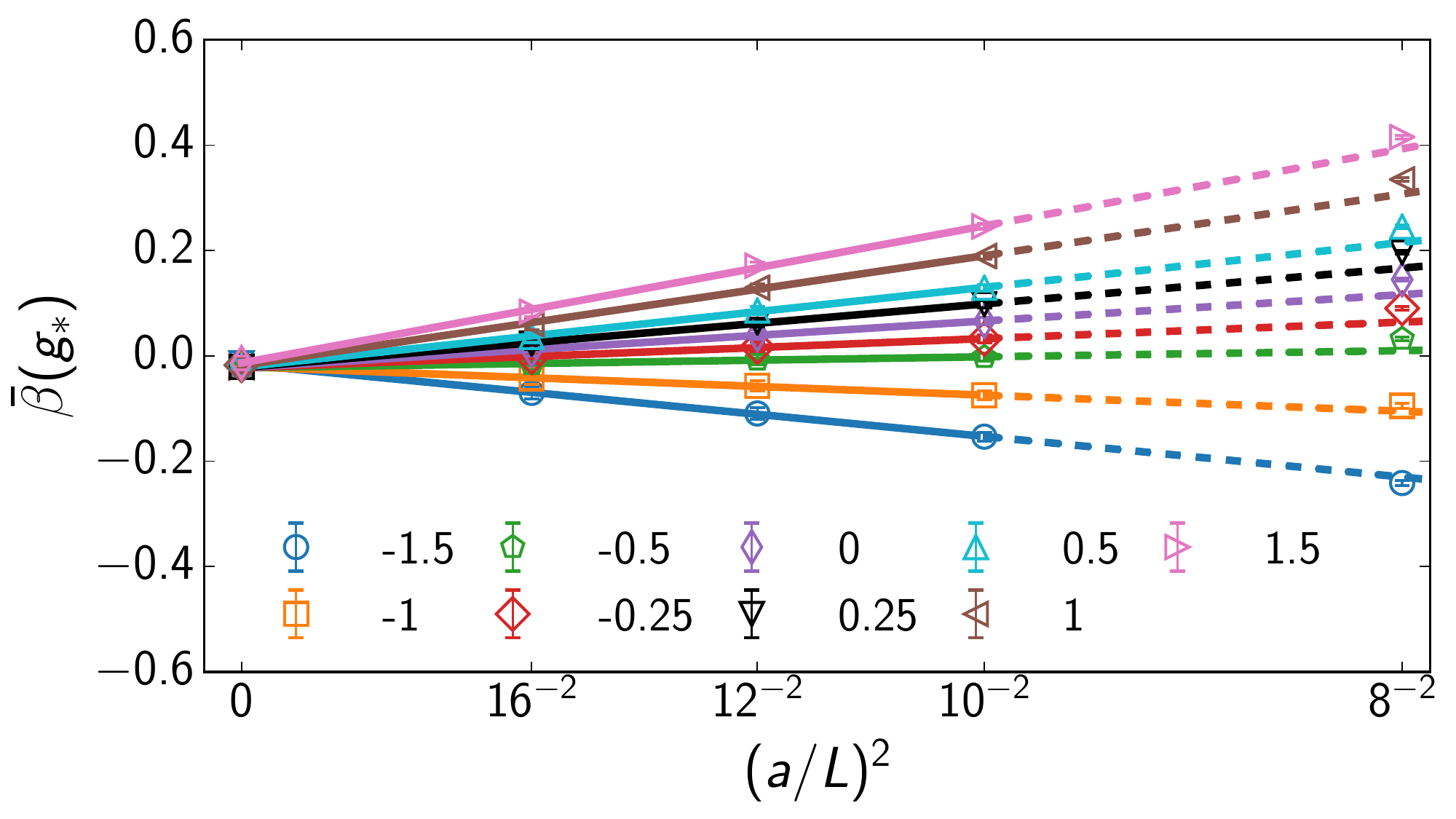}
  \caption[b]{The continuum limit of the $\beta$-function at the IRFP in eight flavor theory,
              for different choices of the discretization mixing parameter $X$ using the
			  quadratic fit.
  }
\label{fig:x8_values_comp}
\end{figure}%

Similarly as in the six flavor case studied above, in Fig.~\ref{fig:x8_values_comp} 
we show the continuum limit of $\bar\beta(g_\ast)$ near the IRFP $g_\ast\sim 2.8$
at different mixing parameters $X$, assuming discretization errors ${\cal O}((a/L)^2)$ in the eight flavor theory. 
From the figure we can see that the continuum limit remains stable with respect to the variations of the parameter $X$, 
but clearly values around $X\sim-0.5$ have reduced $a^2$-effects as the slope is small. 
On the other hand, we do not observe good scaling with the value $X=1.25$ 
as suggested by perturbation theory \cite{Kamata:2016any}.
The quality of the fit is very good near the IRFP, as $\chi^2/\mathrm{d.o.f}$ varies between $0.1$ and $0.5$ depending 
on the mixing parameter $X$.

\begin{figure}[t]
  \includegraphics[width=8.6cm]{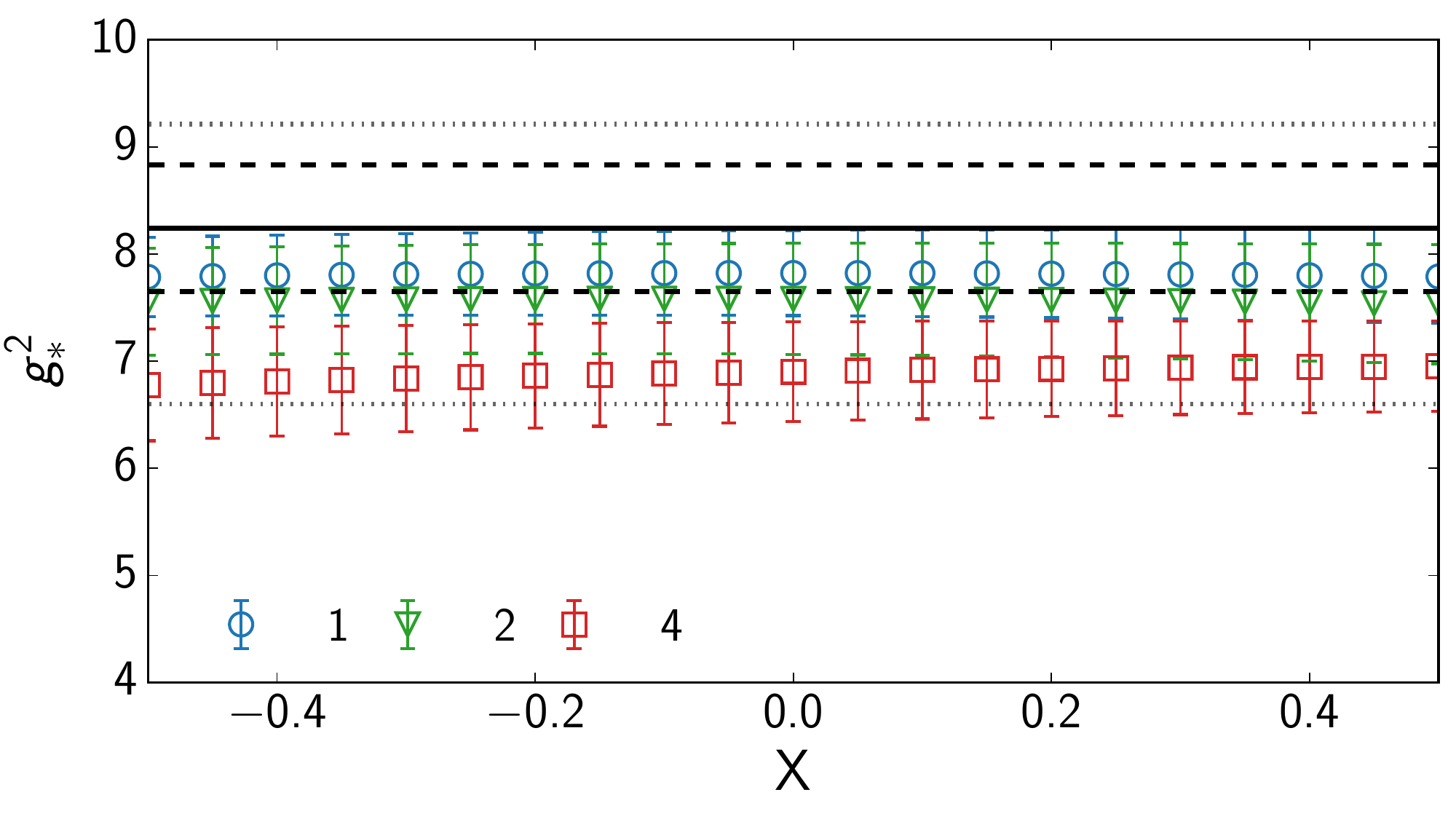}
  \caption[b]{Location of the IRFP in $N_f=8$ theory at different mixing parameter $X$ and different fit ansatz.
              The black lines correspond to the statistical errors and the gray lines give the variance between discretizations
              from~\cite{Leino:2017lpc}.
  }
\label{fig:nf8_mixgpar_aff_g2}
\end{figure}%

In Fig.~\ref{fig:nf8_mixgpar_aff_g2} we show the location of the
IRFP as a function of the mixing parameter $X$ and for different fits.
Compared with the earlier analysis~\cite{Leino:2017lpc},
$g^2_\ast=8.24(59)^{+0.97}_{-1.64}$,
the existence and location of the IRFP does not change when the methods discussed in this paper
are implemented: all interpolation functions give results consistent with the bounds
given by the variance between different discretizations
and shown by the dashed lines in the figure and corresponding to the
second set of errors in the numerical result quoted above.
All the interpolation functions have very small $X$ dependencies, 
and therefore we again choose the $X=0$ as our reference value.

\begin{figure}[t]
  \includegraphics[width=8.6cm]{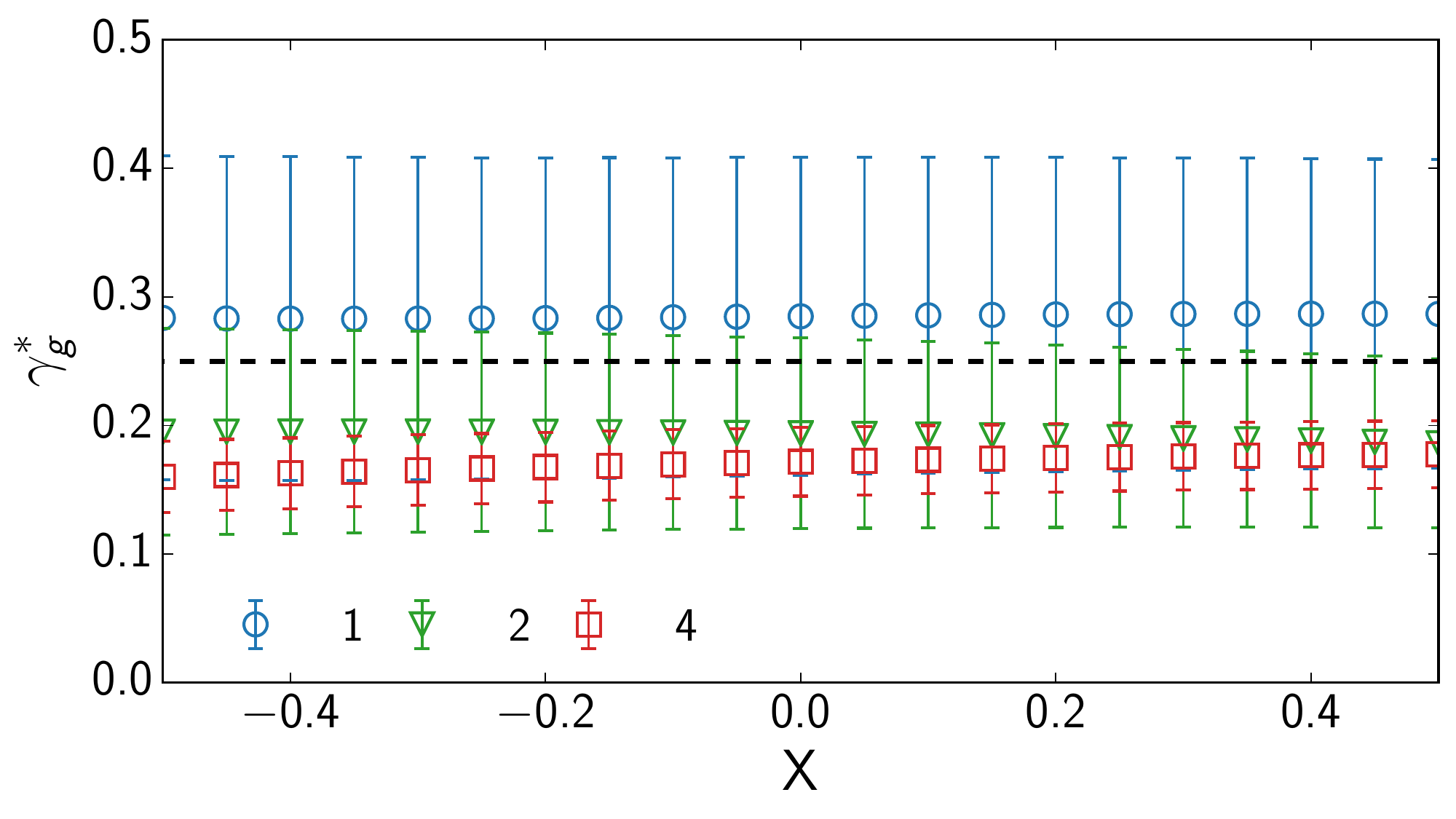}
  \includegraphics[width=8.6cm]{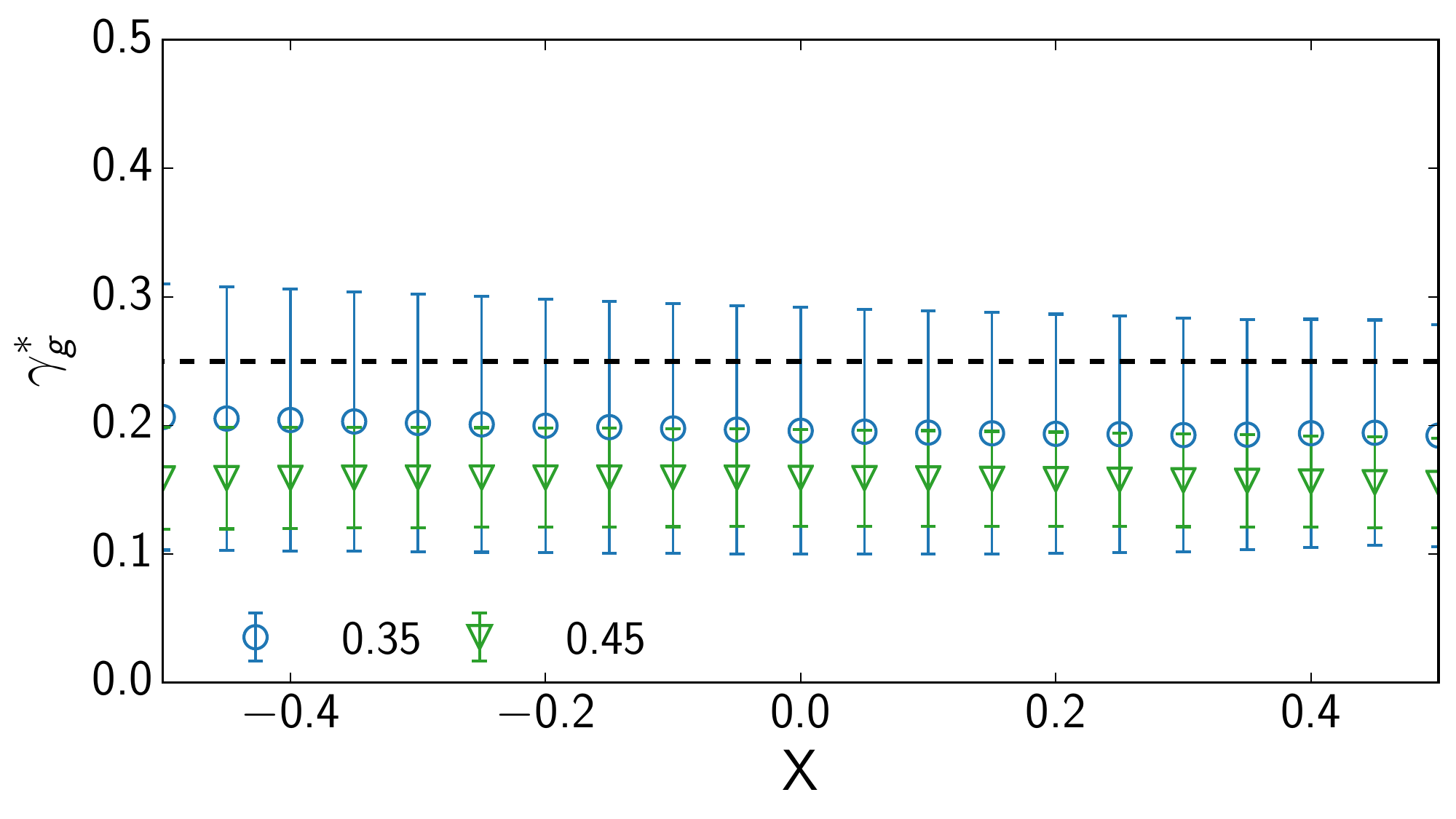}
  \caption[b]{
  			   $\gamma_g^\ast$ for the eight flavor theory, measured at different values of $X$ for
			   {\em top:} different fit ansatz and for {\em bottom:} different values of $c_t$ with quadratic fit.
			   The dashed line shows the scheme independent large-$N_f$ perturbative result $\gamma_g^\ast\approx0.25$
  }
\label{fig:nf8_gammag_variance}
\end{figure}

Similarly, the results for the $\gamma_g^\ast$
measurement are shown in Fig~\ref{fig:nf8_gammag_variance}. The upper panel
shows the results for different fits at $c_t=0.4$. 
The  scheme independent large-$N_f$ perturbative 4-loop result $\gamma^\ast \approx 0.25$~\cite{Ryttov:2017kmx,Ryttov:2017toz} is shown
by the dashed black line.\footnote{The 5 loop result $\gamma^\ast \approx 0.243$ is almost indistinguishable from the 4-loop one.} 
We observe large errors in the result from the linear interpolation. This error is caused by the slight
curve at the fixed point evident in Fig.~\ref{fig:nf8linear}. Because of these large errors,
we quote the quadratic fit as our main result and
measure $\gamma_g^\ast=0.19(8)_{-0.09}^{+0.21}$, 
with the first set of errors being the statistical errors of quadratic fit and the second set of errors give
the variance between different choices of parameter $X$ and interpolation functions.
In order to check the scheme independence of this results, we show in the lower panel of the Fig.~\ref{fig:nf8linear}
the result for different values of scheme parameter $c_t$ for the quadratic fit and $X=0$.
We measure $\gamma_g^\ast=0.16(4)$ for the $c_t=0.45$ case and $\gamma_g^\ast=0.2(1)$ for the $c_t=0.35$ case.
Overall, all the fits when the $X$ is between -0.5 and 0.5 are in agreement with each other and the scheme independent result.

\subsection{Finite-size scaling method}

The results obtained in the previous subsection rely on the $a^2$-scaling of the lattice observables.
Near the IRFP this scaling can be modified by nontrivial scaling exponents.
If these scaling exponents remain small near the infrared fixed point, we
can assume that the power counting argument holds and cutoff effects dominated by dimension 6 operators decrease
with the power of lattice spacing $a$. This can modify the $a^2$-scaling,
which relies on Symanzik improvements around the Gaussian UV fixed point.
Since we checked our continuum limit with multiple different $a^2$ scalings (by varying $X$),
and since the results hold between discretizations and varying $c_t$, we argue that the scaling violation is small
and the continuum limit is robust.

In order to check the consistency of the $a^2$-scaling in the continuum limit, we also measure
the leading irrelevant exponent using a finite-size scaling method
developed in
Refs.~\cite{Appelquist:2009ty,DeGrand:2009mt,Lin:2015zpa,Hasenfratz:2016dou}.
In the close proximity of the IRFP, by integrating the $\beta$-function,
we obtain a scaling relation between lattices of size $L_\mathrm{ref}$ and $L$~\cite{Lin:2015zpa}:
\begin{equation}
\gGF^2(\beta,L) - g_\ast^2 =
\left[ \gGF^2(\beta,L_\mathrm{ref})-g_\ast^2\right]\left(\frac{L_\mathrm{ref}}{L}\right)^{\gamma_g^\ast}\,.
\label{eq:ramosgamma}
\end{equation}
This equation relies on the evolution of the coupling towards the fixed point as the lattice size is increased
from $L_\mathrm{ref}$ to $L$. Hence, it cannot be used exactly at the IRFP where there is no evolution and
$\gGF^2-g_\ast\sim0$.

\begin{figure}[t]
  \includegraphics[width=8.6cm]{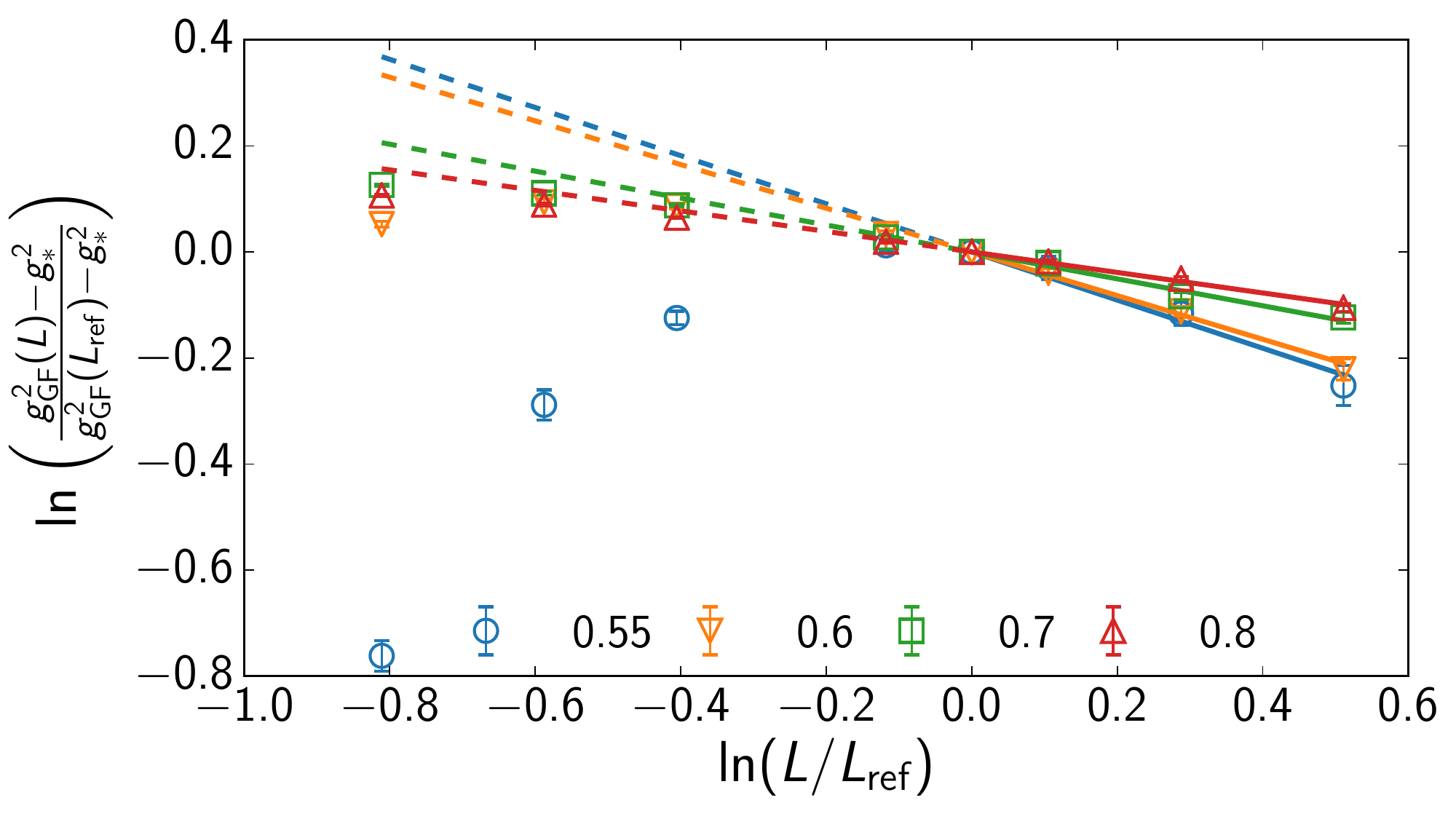}
  \includegraphics[width=8.6cm]{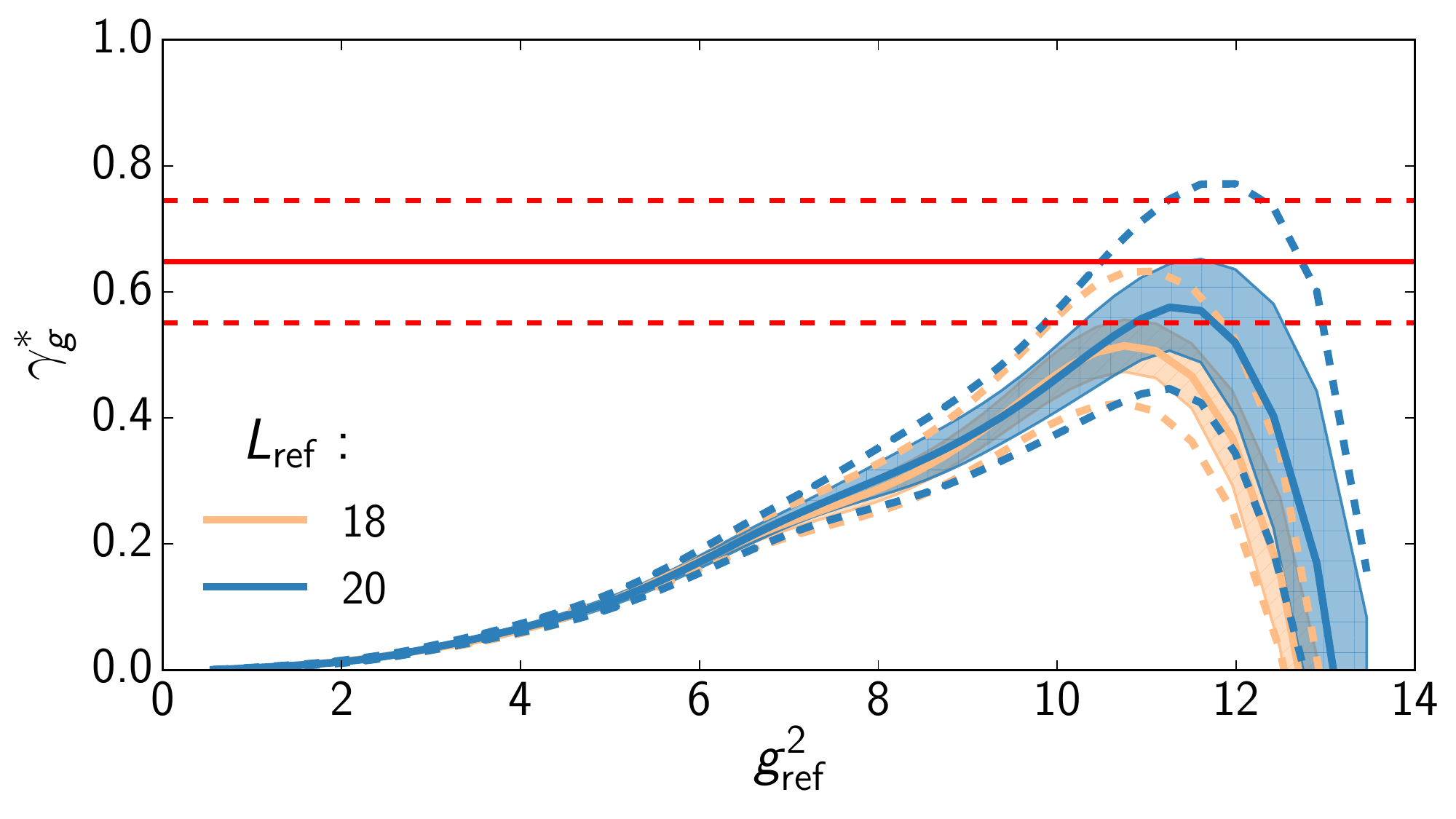}
  \caption[b]{
		  {\em top:} Fit to function~\eqref{eq:ramosgamma} for measured couplings $\gGF(\beta_L,L)$ 
		  with $\tau_0$-correction at $\beta_L=  0.55 \ldots 0.8$ using $L_\mathrm{ref}/a=18$.
		  {\em Bottom:} The $\gamma_g^\ast$ with the finite size scaling method for $N_f=6$.
          The red lines indicate the result from the slope of the $\beta$-function, $\gamma_g^\ast=0.648(97)_{-0.1}^{+0.16}$.
  }
\label{fig:nf6_ramos_method_fromnf6paper}
\end{figure}

Again, we start with the analysis of $N_f=6$ theory.
We applied this method in Ref.~\cite{Leino:2017hgm} for the $N_f=6$ model and 
obtained the fit presented in Fig.~\ref{fig:nf6_ramos_method_fromnf6paper}.
The figure shows the $L$ dependence of the fit to function~\eqref{eq:ramosgamma} and 
the final measurement of $\gamma_g^\ast$ in six flavor theory for two different values of $L_{\rm{ref}}/a=18$ and 20.
We use a polynomial interpolation to the $\tau_0$-corrected measurements and
choose the IRFP to be at the measured value
$g^2_\ast=14.5(3)^{+0.41}_{-1.38}$.
The lattice sizes are varied between $L_\mathrm{ref}$ and $30$.
The dashed lines around the shaded bands show the
variation of the result when $g^2_\ast$  is varied within
its statistical errors.

Since the method breaks down at the fixed point, where $\gGF^2-g_\ast\sim0$,
we quote the maximum value as the most probable $\gamma_g^\ast$ measurement with this method. 
This is not reliable measurement of $\gamma_g^\ast$, but offers a consistency check for our earlier result 
from the slope of the $\beta$-function.

\begin{figure}[t]
  \includegraphics[width=8.6cm]{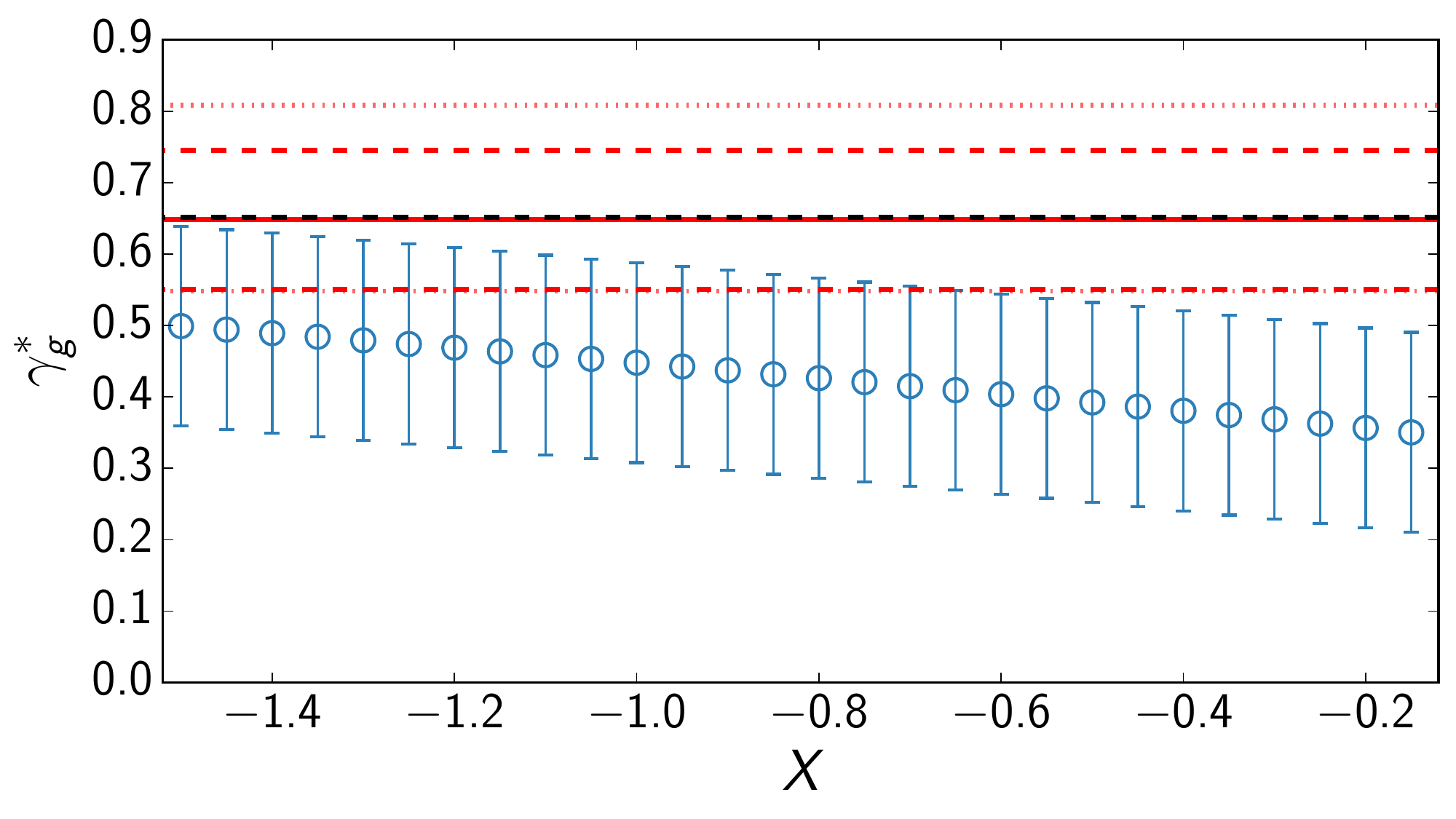}
  \caption[b]{The the dependence on $X$ of the maximum value of $\gamma_g^\ast$ obtained with the finite scaling method.
  $L_\mathrm{ref}/a=16$.}
\label{fig:nf6_ramos_method_xdep}
\end{figure}

Furthermore, this method assumes vanishing discretization artifacts
and thus it can only be used in the region of parameter space that is close to continuum (i.e. large $L$)
and where the lattice artifacts are small.
In order to check the dependence on the lattice artifacts, we redo the fit to Eq.~\eqref{eq:ramosgamma}
with different values of the mixing parameter $X$. We take the maximum value to be the most probable measurement
of the $\gamma_g^\ast$ and show our results in the Fig.~\ref{fig:nf6_ramos_method_xdep}.
Here the open symbols show the maximum value of $\gamma_g^\ast$ at each $X$ obtained with the finite size scaling method.
We observe that this method is indeed sensitive to the $a^2$-effects.
Not only does the measurement of $\gamma_g^\ast$ depend on the value of
$X$, but we cannot even get a non-zero result with unimproved data $X>0.1$.
The measurements in the region $X<-0.5$, with small $a^2$-scaling, we get results that agree with the $\tau_0$-corrected result
in Fig.~\ref{fig:nf6_ramos_method_fromnf6paper}.
Overall, for $X<-0.5$ this analysis yields results consistent with the earlier analysis~\cite{Leino:2017hgm} 
and also with the theoretical scheme independent result~\cite{Ryttov:2017toz}.

\begin{figure}[t]
  \includegraphics[width=8.6cm]{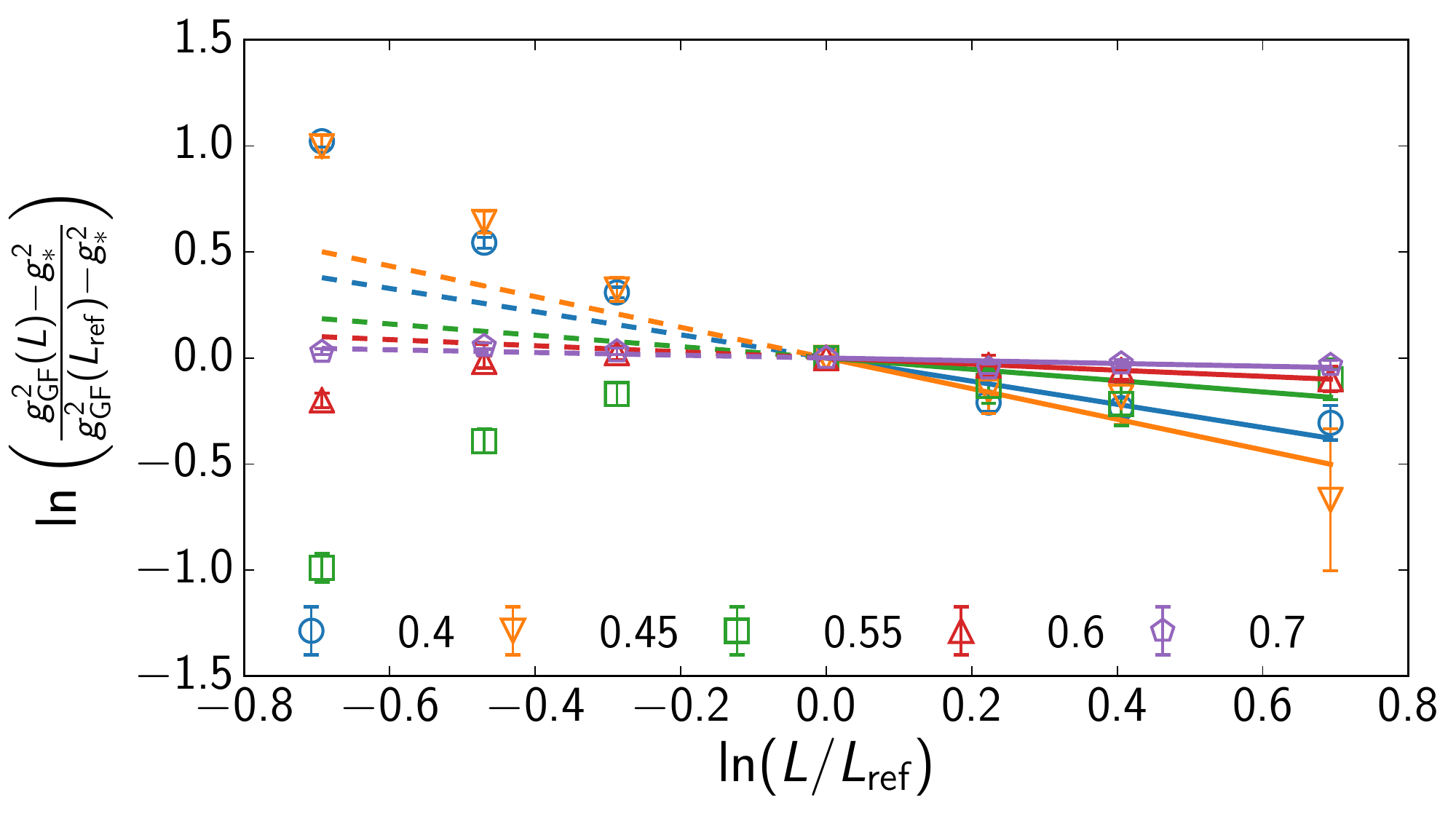}
  \caption[b]{Fit to function~\eqref{eq:ramosgamma} for measured couplings $\gGF^2$ with $\tau_0$-correction at
  $\beta=0.4\ldots0.7$ at $c_t=0.4$ using $L_\mathrm{ref}/a=16$.
  }
\label{fig:nf8_ramos_method_raw}
\end{figure}

Finally, we show similar results for the $N_f=8$ theory. 
Since this method is very sensitive on $a^2$-effects and the $\tau_0$-correction most consistently improves the 
data for the full range of measured couplings, we will do the fit to Eq.~\eqref{eq:ramosgamma}
using the results from~\cite{Leino:2017lpc} with $\tau_0$-correction,
$c_t=0.4$, $L_{\rm{ref}}/a=16$ and $\beta$ between 0.4 and 0.7. This fit is shown in Fig.~\ref{fig:nf8_ramos_method_raw}.


\begin{figure}[t]
  \includegraphics[width=8.6cm]{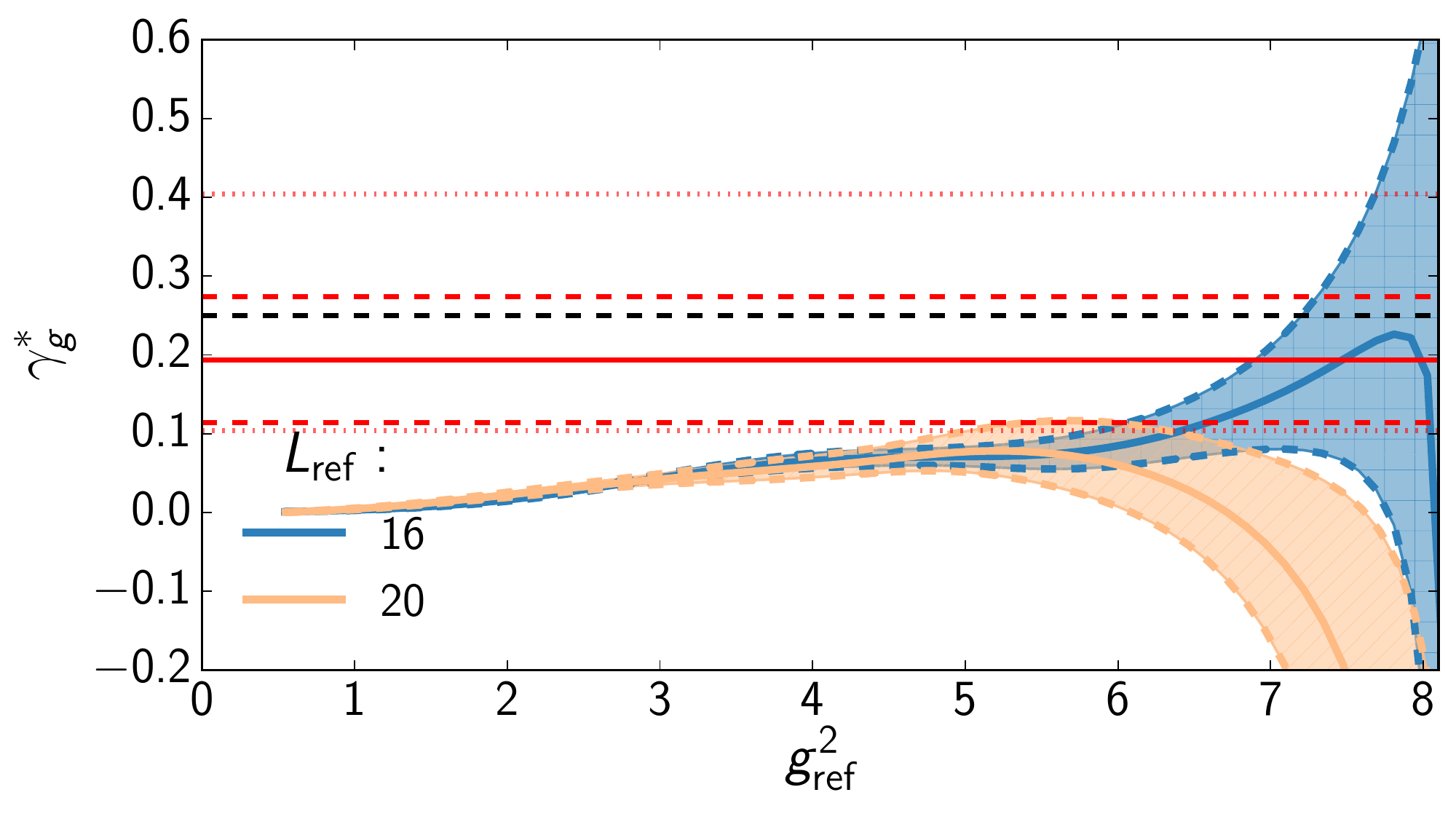}
  \caption[b]{The $\gamma_g^\ast$ with the finite size scaling method for $N_f=8$. The red lines indicate the result from the slope, $\gamma_g^\ast=0.28(12)$, and the black line shows the scheme invariant estimate.
  }
\label{fig:nf8_ramos_method}
\end{figure}%
\begin{figure}[h]
  \includegraphics[width=8.6cm]{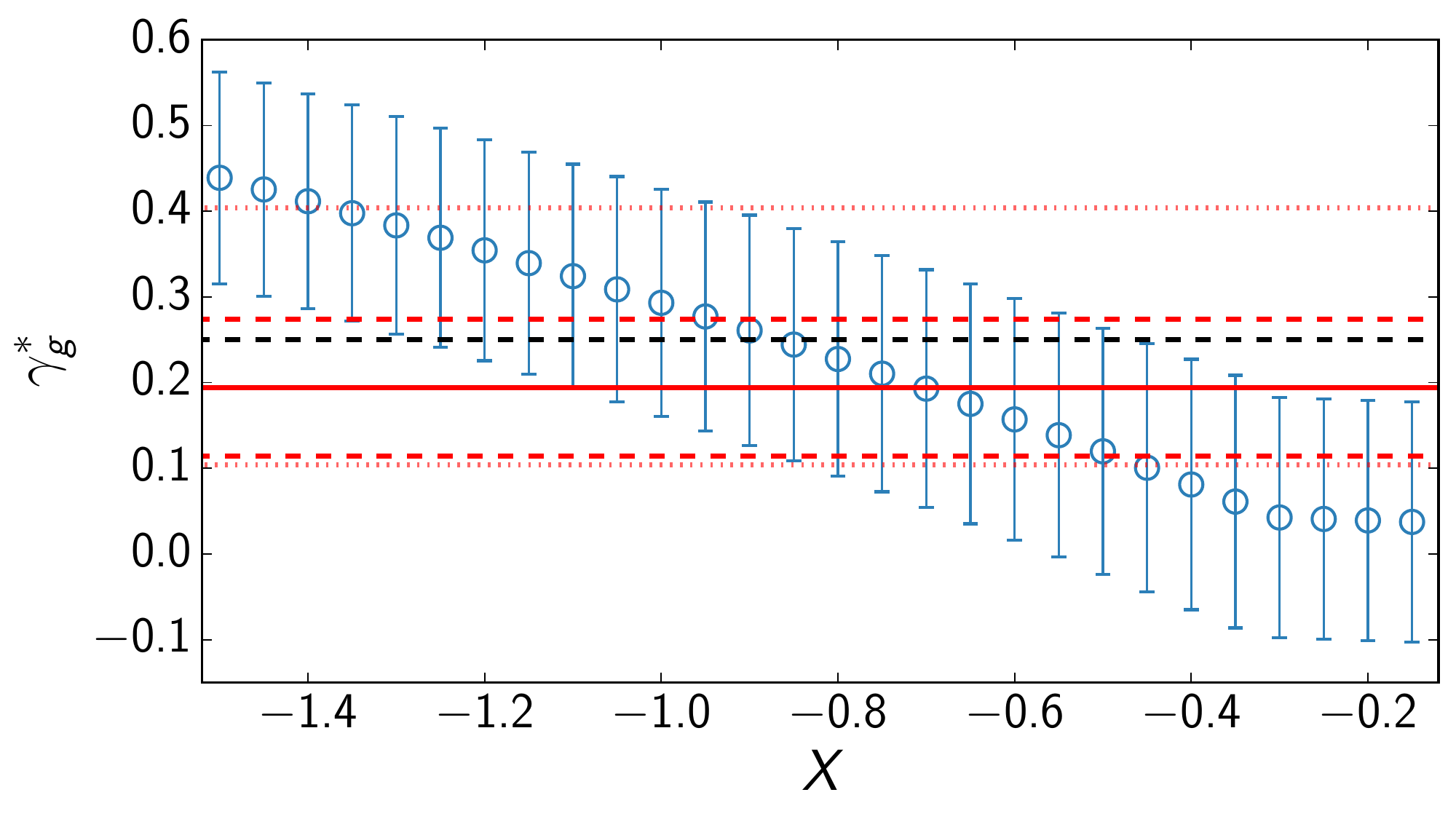}
  \caption[b]{The the dependence on $X$ of the maximum value of $\gamma_g^\ast$ obtained with the finite scaling method.
  $L_\mathrm{ref}/a=16$.
  }
\label{fig:nf8_ramos_method_xdep}
\end{figure}%

In Fig.\ref{fig:nf8_ramos_method} we show the value of $\gamma_g^\ast$ obtained
by the finite size scaling method in the eight flavor theory.
In the figure the horizontal red lines correspond to the value $\gamma_g^\ast=0.28(12)$ obtained
from the slope of the $\beta$-function at the fixed point 
and the black line corresponds to the scheme independent result $\gamma_g^\ast=0.25$. To obtain
the finite size scaling result we use $L_{\rm{ref}}=16$ or 20 and vary the lattice sizes between $L_\mathrm{ref}$ and $32$. 
Rational interpolation of the $\tau_0$-corrected measurements
is used and the measured fixed point value $g^2_\ast=8.24(59)^{+0.97}_{-1.64}$.
We observe that with $L_{\rm{ref}}/a=16$ this method gives results consistent with the earlier measurement
with the slope of the $\beta$-function, while the $L_{\rm{ref}}/a=18$ gives a result slightly smaller.

Again, we check the dependence on $a^2$-effects by redoing the analysis without the $\tau_0$-correction and
varying the mixing parameter $X$. The $X$-dependence of $\gamma_g^\ast$ is shown in Fig.\ref{fig:nf8_ramos_method_xdep}.
We observe that for the eight flavor case, the $a^2$ dependence is even stronger than for the six flavor case,
which renders this method unreliable. With the mixing parameter within $X=-0.2\ldots-1.5$ we observe results
consistent with the $\gamma_g^\ast$ measured from the slope of $\beta$-function.


\begin{table}
\begin{ruledtabular}
\begin{tabular}{cccccccc}
 $N_f$ & $\gamma_{2,\mathrm{RS}}$ & $\gamma_{3,\mathrm{RS}}$ & $\gamma_{4,\mathrm{RS}}$ & 
 $\gamma_{5,\mathrm{RS}}$ & $\gamma_{2}$ & $\gamma_{3}$ & $\gamma_{4}$ \\
\hline
6 & 0.499 & 0.957 & 0.734 & 0.6515 & 6.06 & 1.62 & 0.974 \\
8 & 0.180 & 0.279 & 0.250 & 0.243  & 0.4 & 0.3181 & 0.2997 \\
\end{tabular}
\end{ruledtabular}
\caption{\label{tab:chi2}
Perturbative values for $\gamma_g^\ast$ at 2- to 5-loop level, denoted with $\gamma_2 \ldots \gamma_5$.  The subscript RS refers to the scheme independent large $N_f$ calculation by Ryttov and Shrock~\cite{Ryttov:2017kmx,Ryttov:2017toz},
and the results without RS are $\MSb$ results by Herzog et al. \cite{Herzog:2017ohr}. At 5-loop level the $\MSb$ result does not have an IRFP.
}
\end{table}%
\FloatBarrier

\section{Conclusions and Outlook}
\label{sec:checkout}
We have studied the properties of the IRFP of SU(2) lattice gauge theory
with 6 or 8 fermions in the fundamental representation.
The existence of IRFP in these theories has been established in earlier
work~\cite{Leino:2017lpc,Leino:2017hgm},
and in this paper we focused on determination of the leading irrelevant
critical exponent, $\gamma_g^\ast$ in these two theories.
For the first time, we obtain in the eight flavor theory the result
$\gamma_g^\ast=0.19(8)_{-0.09}^{+0.21}$.
This result is compatible with the scheme independent large $N_f$ perturbative result $\approx 0.243$ 
obtained in~\cite{Ryttov:2017kmx,Ryttov:2017toz}.
For a detailed comparison of the scheme independent large-$N_f$ and $\MSb$ results
we refer the reader to Table~\ref{tab:chi2}.
We also studied the robustness of the results with respect to different interpolations used in the analysis.

Furthermore, we have shown that the methodology we have applied here is consistent with the earlier analysis of the six-flavor theory.
In this paper we obtained the result $\gamma_g^\ast=0.66(4)^{+0.25}_{-0.13}$ using a quadratic fit to the $\beta$-function near the IRFP.
This result was shown to be very stable with respect to the discretization mixing parameter in the definition of the gradient flow.
The result is also consistent with the earlier result $\gamma_g^\ast=0.648(97)_{-0.1}^{+0.16}$ in~\cite{Leino:2017hgm}
and hence also with the analytic results of~\cite{Ryttov:2017kmx,Ryttov:2017toz}.
Again, the perturbative results are presented in Table~\ref{tab:chi2}.

Our results indicate the emergence of a consistent picture of strong coupling
features of SU(2) gauge theory inside the conformal window.

\FloatBarrier

\acknowledgments{
This work is supported by the Academy of Finland grants 308791 and 310130.
V.L acknowledges support from the Jenny and Antti Wihuri foundation.
}

\bibliography{su2_nf6.bib}{}
\bibliographystyle{apsrev4-1.bst}

\FloatBarrier

\end{document}